\documentclass[sigplan,10pt, nonACM]{acmart}
\renewcommand\footnotetextcopyrightpermission[1]{}
\usepackage[normalem]{ulem} 
\usepackage{amssymb,amsmath} 
\usepackage{openbib} 
\usepackage{multirow,tabularx}
\usepackage{fancyhdr} 
\usepackage{comment} 
\usepackage{subfig} 
\usepackage{float} 
\usepackage{xspace} 
\usepackage{footmisc} 
\usepackage{listings} 
\usepackage{tabularx,tabulary,booktabs} 
\usepackage{afterpage} 
\usepackage{soul} 
\usepackage{graphicx} 
\usepackage{color}
\usepackage{enumitem}

\newcommand{\des}{\textit{Maya}\xspace}
\newcommand{\apps}{\textit{elastic}\xspace}

\newcommand{\rpp}[1]{#1}

\begin{document}
\settopmatter{printfolios=true}
\settopmatter{printacmref=false}

\title{Maya: Falsifying Power Sidechannels with Dynamic Control}
\author{Raghavendra Pradyumna Pothukuchi, Sweta Yamini Pothukuchi, Petros Voulgaris, Alexander Schwing, Josep Torrellas}
\affiliation{University of Illinois at Urbana-Champaign}

\begin{abstract}
The security of computers is at risk because of information leaking through  physical outputs such as power, temperature, or electromagnetic (EM) emissions. Attackers can use advanced signal measurement and analysis to recover sensitive data from these sidechannels.

To address this problem, this paper presents {\em \des},
a simple and effective solution against power
side-channels. The idea is to {\em re-shape} the power dissipated by an application in an application-transparent manner using control theory techniques -- preventing attackers from learning any information.
With control theory, a controller can reliably 
keep power close
to a desired target value even when runtime conditions change
unpredictably. Then, by changing these targets 
intelligently, power can be made to appear in any desired
form, appearing to carry activity information which, in
reality, is unrelated to the application.
\des can be implemented in privileged
software or in simple hardware.  In this paper, we 
implement \des on two multiprocessor machines using Operating System (OS)
threads, and show its effectiveness and ease of deployment.

\end{abstract}

\maketitle

\section{Introduction}
\label{intro}

There is an urgent need to secure computer systems against the growing number of cyberattack surfaces. An important class of these attacks is that which utilizes the physical outputs of a system, such as its power, temperature or electromagnetic (EM) emissions. These outputs are correlated with  system activity and can be exploited by attackers to recover sensitive information~\cite{mobilePowerside,powerspy,batteryAttack,Kocher2011,sideChannelRetrospect}.
 
Many computing systems ranging from mobile devices to multicore servers in the cloud are vulnerable to physical side channel attacks~\cite{mobilePowerside,powerspy,batteryAttack,powerSide,usbPowerSide,thermalMulticoreCovert}. With advanced signal measurement and analysis, attackers can identify many details like keystrokes, password lengths~\cite{mobilePowerside}, personal data like  location, browser and camera activity~\cite{batteryAttack,usbPowerSide,powerspy}, and the bits of encryption keys~\cite{Kocher2011}. 

Many defenses against physical sidechannels have been proposed which aim to keep the physical signals constant or noisy~\cite{constPowerLogic,maskDesEnergy,yang2005power,circuitPowerDefense,Kocher2011,dvfsDefense,randomDvsDfs}. However, all these techniques require new hardware and, hence, existing systems in
the field are left vulnerable. Trusted execution environments like Intel SGX~\cite{sgx} or ARM TrustZone~\cite{trustzone} cannot ``contain'' physical signals and are ineffective to stop information leaking through them~\cite{dvfsArmCovert,trustzonePowerside,thermalMulticoreCovert}.  

To address this problem,
we seek a solution that is simple to implement and is effective against power
side-channels. The idea is to rely on privileged software or simple hardware to distort, in an application-transparent manner, the power dissipated by an application --- so that the attacker cannot glean any information.
Obfuscating power also removes leakage through temperature and EM signals, since they are  directly related to the computer's power~\cite{emPower,trustzonePowerside, thermalMulticoreCovert}. Such a defense can prevent exploits that analyze application behavior at the scale of several milliseconds or longer, such as those that infer what applications are running, what data is used in the camera or browser, or what keystrokes occur~\cite{batteryAttack,mobilePowerside}.

A first challenge in building an easily deployable  power-shaping defense
is the lack of configurable system inputs that can effectively change power. Dynamic Voltage and Frequency Scaling (DVFS) is an input supported by nearly all mainstream processors~\cite{sandybridge_power,skylakePower, power8Dvfs,zeppelin,cortexA7,cortexA15}. However, DVFS levels are only a few, and the achievable range of power values depends on the application ---  compute-intensive phases have higher values of power and show bigger changes with DVFS, while it is the opposite for memory-bound applications.  
Injecting idle cycles in the system~\cite{powerclamp,linaroidle} is another technique,
but it can only reduce power and not increase it. 

The second and more important challenge is to develop an algorithm that reshapes
power to effectively eliminate any information leakage. This is hard because applications vary widely in their activity profile and in how they respond to system inputs. Attempts to maintain constant power, insert noise into power signals, or simply randomize DVFS levels have been unsuccessful~\cite{batteryAttack,shapeSignal,powerSide}. These techniques only tend to add noise, but do not mask application activity~\cite{shapeSignal,powerSide}.

In this work, we propose {\em \des}, a defense architecture 
that intelligently re-shapes a computer's power in an 
application-transparent manner, so that the attacker
cannot extract application information. To achieve this, 
\des uses a Multiple Input Multiple Output (MIMO) controller from  control theory~\cite{mimoText}. 
This controller can reliably keep power close to a target power level even when runtime conditions change unpredictably. 
By changing these targets intelligently, 
power can be made to appear in 
any desired form, appearing to carry activity information which,
in fact, is  unrelated to the application.


\des can be implemented in privileged
software or in simple hardware. In this paper, 
we implement \des on two multicore machines using OS threads.
The contributions of this work are:
\vspace{-1mm}

\begin{enumerate}[leftmargin=4mm]
    \item \des, a new defense system against power side-channels through power 
  re-shaping.

    \item An implementation of \des using only privileged software. To the best of our knowledge, this is the first software defense against power side-channels that is application-transparent.

    \item The first application of MIMO control theory to side-channel defense.

    \item An analysis of power-shaping properties necessary to mask information leakage. 

    \item An evaluation of \des using  machine learning-based attacks on two different real machines.
\end{enumerate}

\section{Background}
\label{back}

\subsection{Physical Side-Channels}
\label{sub_physical}
\label{sub_overview}

Physical side-channels such as 
power, temperature, and electromagnetic (EM) emissions 
carry significant information about the execution. For example,
these side channels
have been used to 
infer the characters typed by a smartphone user~\cite{batteryAttack},
to identify the running application, login requests,
and the length of passwords on a commodity 
smartphone~\cite{mobilePowerside}, and even to
recover  the full encryption key from a cryptosystem~\cite{kocher1999differential}. 

All power analysis attacks rely on the principle that the 
dynamic power of a computing system is directly proportional 
to the switching activity of the hardware. Since this activity 
varies across instructions, groups of instructions, and application tasks, they all leave distinct power traces~\cite{armInstEnergy,x86InsEnergy,batteryAttack,mobilePowerside}. By analyzing these power traces, many details 
about the execution can be deduced. 
Temperature and EM emissions are also directly related to a computer's power consumption and techniques to analyze them are similar~\cite{emPower,trustzonePowerside, thermalMulticoreCovert}.

Physical side-channels can be sensed through special measurement devices~\cite{Kocher2011}, unprivileged hardware and OS counters~\cite{sandybridge_power,androidBattery}, public amenities like USB charging stations~\cite{usbPowerSide}, malicious smart batteries~\cite{batteryAttack} or remote antennas that measure EM emissions~\cite{genkinEm,GenkinLaptop}. In cloud systems, an application can use the thermal coupling between cores to infer the temperature profile of a co-located application~\cite{thermalMulticoreCovert}. 

After measuring the signals, attackers search for patterns in the signal over time like phase behavior and peak locations, and its frequency
spectrum after a Fourier transform. 
This can be done through Simple Power Analysis (SPA), which
uses a single trace~\cite{batteryAttack,genkinEm,mobilePowerside},
or Differential Power Analysis (DPA), which examines the
statistical differences between thousands of 
traces~\cite{Kocher2011,powerSide}.  

The timescale over which the signals are analyzed is determined by the information that attackers seek. For cryptographic keys, it is necessary to record and analyze signals over a few microseconds or faster~\cite{Kocher2011}. For higher-level information like 
the details of the running applications, keystrokes, browser data, and personal information, signals are analyzed over timescales of milliseconds or more~\cite{usbPowerSide,batteryAttack,mobilePowerside}.
The latter are the timescales that this paper focuses on.

\subsection{Control Theory Techniques}
\label{sub_robust}

\begin{figure}[h]
\centering
\includegraphics[width=\columnwidth]{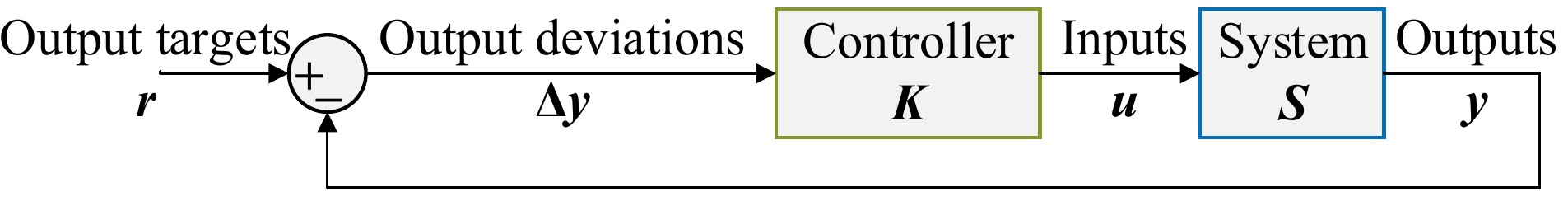}
\vspace{-5mm}
\caption{Control loop.}
\vspace{-3mm}
\label{fig_Loop}
\end{figure}

Using control theory~\cite{mimoText}, we can
design a controller $K$ that manages a system $S$ (i.e., a computer)
as shown in Figure~\ref{fig_Loop}. The system has outputs $y$ (e.g., 
the power consumed) and  configurable inputs $u$ (e.g., the DVFS level). 
The outputs must be kept close to the output targets $r$. 
The controller reads the 
deviations of the outputs from their targets ($\Delta y =r -y$), and 
sets the inputs.

The controller is a state machine characterized by a state vector, $x(T)$, 
which evolves over time $T$. It advances its state to $x(T+1)$ and
generates the system inputs $u(T)$ by reading the output deviations $\Delta y(T)$: 
\begin{equation}
 \begin{aligned}
        x(T+1) &= A\times x(T) + B\times \Delta y(T)\\
        u(T) &= C\times x(T) + D\times \Delta y(T)
       \end{aligned}
 \quad x(0) = 0
 \label{eq:control}
\end{equation}
\noindent
where $A$, $B$, $C$, and $D$ are matrices that encode the controller. 
The most useful controllers are those that actuate on
Multiple Inputs and sense Multiple Outputs (MIMO) at the same time.

Designers can specify multiple parameters in the control system~\cite{mimoText}.
They include the maximum bounds on the deviations of the outputs from their
targets, the differences from design conditions or unmodeled effects that the controller must be tolerant to (i.e., the uncertainty guardband), and the relative
priority of changing the different inputs, if there is a choice
(i.e., the input weights). With these
parameters, controller design is automated~\cite{robustMATLAB}.

\section{Threat Model}
\label{sub:thread}

We assume that attackers try to compromise the victim's security through power measurements. They measure power 
using off-the-shelf sensors present in the victim machine like hardware counters or OS APIs~\cite{sandybridge_power,androidBattery}. Such
sensors are only reliable at the time granularity of several milliseconds. Attackers could use alternative measurement strategies at this timescale like malicious USB charging booths~\cite{usbPowerSide}, compromised batteries~\cite{batteryAttack}, 
thermal measurements~\cite{thermalMulticoreCovert}, or cheap power-meters when physical access is possible~\cite{wei19host}. 

We exclude power analysis attacks that search for patterns at a fine time granularity with special hardware support such as oscilloscopes~\cite{kocher1999differential} or antennas~\cite{genkinEm,GenkinLaptop}.
While such attacks 
are powerful enough to attack cryptographic algorithms~\cite{Kocher2011}, they
are harder to mount and need more expensive equipment. Even with fine-grain measurement, information about events like keystrokes can be analyzed only at the millisecond timescale~\cite{batteryAttack,mobilePowerside}.

We assume that attackers can know the algorithm used by \des to reshape the computer's power. They can run the algorithm and see its impact on the time-domain and frequency-domain behavior of applications. Using these observations, they can develop machine learning models to adapt to the defense and try defeating it.

Finally, we assume that the hardware or privileged software that implements the control system to reshape the power trace is uncompromised. In a software implementation, the OS scheduler and DVFS interfaces are assumed to be uncompromised.

\section{Falsifying Power Side-channels with Control}
\label{goal}




We propose that a computer system 
defend itself against power side-channel attacks by
distorting its pattern of power consumption. 
Unfortunately,
this is hard to perform successfully. Simple distortions like adding noise to power signals can be removed by attackers using signal processing techniques.
This is especially the case
if, as we assume in this paper, the attacker knows the general algorithm
that the defense uses to distort the signal. Indeed, past approaches have been unable to provide a solution to this problem. In this paper, we propose a new approach.
In the following, we describe
the rationale behind the approach, present the high-level architecture, 
and discuss how 
to generate the distortion to falsify information leakage through power signals.

\subsection{Why Use Control Theory Techniques?}
\label{why}

\begin{figure}[h]
\centering
\subfloat[Original]{
\includegraphics[width=0.4\columnwidth]{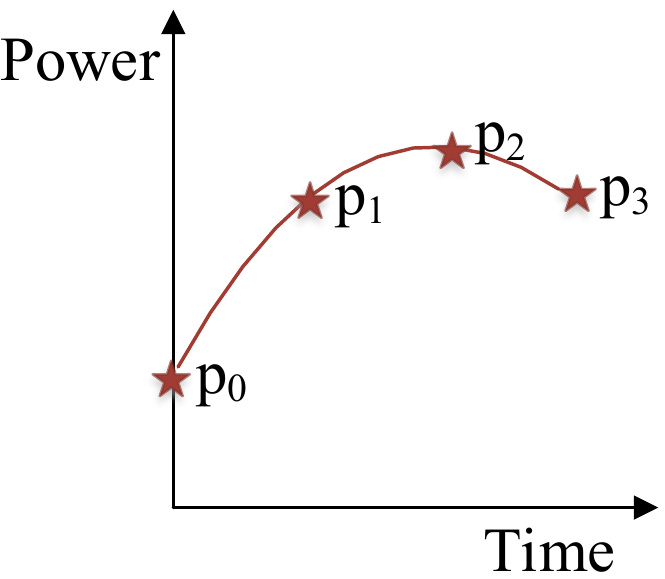}
\label{subfig_o}
}
\subfloat[Distorted]{
\includegraphics[width=0.4\columnwidth]{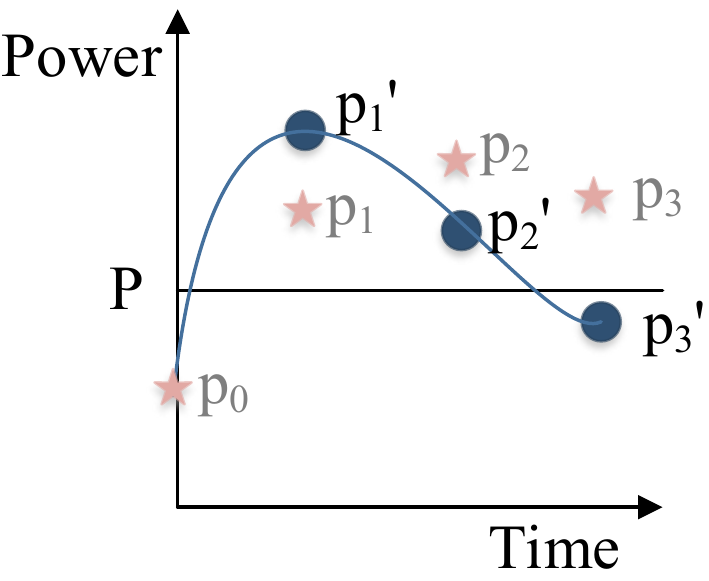}
\label{subfig_d}
}
\caption{Example of the power trace for an application.}
\label{trace}
\end{figure}

To understand why control theory techniques are necessary at shaping power, consider the following scenario. 
We measure the power consumed by an application at fixed timesteps to record a trace, as shown in Figure~\ref{subfig_o}.
To prevent information leakage, we must distort the power trace into a different uncorrelated shape.

Assume that the system has a mechanism to increase the power consumed, and another to reduce it. 
In this paper, these mechanisms are the ability to run a {\em Balloon} thread and an idle thread, respectively,
for a chosen amount of time. A balloon thread is one that performs power-consuming operations (e.g., floating-point operations) in a tight loop, and an idle thread forces the processor into an idle state. 

One way to mislead the attacker is to keep the power consumption at another level $P$ (Figure~\ref{subfig_d}).
To achieve this, we can measure the difference between $P$ and the actual power $p_i$ at each timestep and schedule a balloon thread if $P-p_i > 0$ or an idle thread otherwise. Unfortunately, this approach is too simplistic to be effective.
First, it ignores how the application's power changes intrinsically. Second, achieving the power $P$ with this application may require a combination of both the idle and balloon threads. When only a balloon thread is scheduled at the 0$^{th}$ timestep based on $P-p_0$, the power in the 1$^{st}$ timestep would be $p_1'$ rather than our target $P$. 


If this poor control algorithm is repeatedly applied, it will always miss the target and we obtain the trace in Figure~\ref{subfig_d} where the measured power is not close to the target, and in addition, has enough features of the original trace.


An approach with control theory is able to get much closer to the target power level. This is because the controller makes more informed power changes at every interval and can set multiple inputs for accurate control. 
To understand why, we rephrase the equations of controller operation (Equation~\ref{eq:control}) slightly: 
\begin{equation}
 \begin{aligned}
        State(T+1) &= A\times State(T) + B\times Error(T)\\
        Action(T) &= C\times State(T) + D\times Error(T)
       \end{aligned}
 \label{eq:control2}
\end{equation}
The second equation shows that the action taken at time $T$ (in our case, scheduling the balloon and idle threads) is a function of the tracking error observed at time $T$ (in our case, $P$-$p_0$) \textit{and} the controller's state. The state is a summary of the controller's experience in regulating the application. The new state used in the next timestep is determined by the current state and error serving as an ``accumulated experience'' to smoothly and quickly reach the target. 

Further, the controller's actions and state evolution are influenced by the matrices
$A$, $B$, $C$, and $D$, which were generated when the controller was designed. This design includes running a {\em training set} of applications
while scheduling the balloon and idle threads and measuring the resulting power changes.
Consequently, these matrices embed the intrinsic behavior of the applications under these conditions. 

Overall, with a control theory controller the measured power trace will be much closer to the target signal. 
If the target signal is chosen appropriately, the attacker can longer recover the application information.

%
%

%

\subsection{High-Level Architecture}
\label{high-level}

\begin{figure}[h]
\centering
\includegraphics[width=\columnwidth]{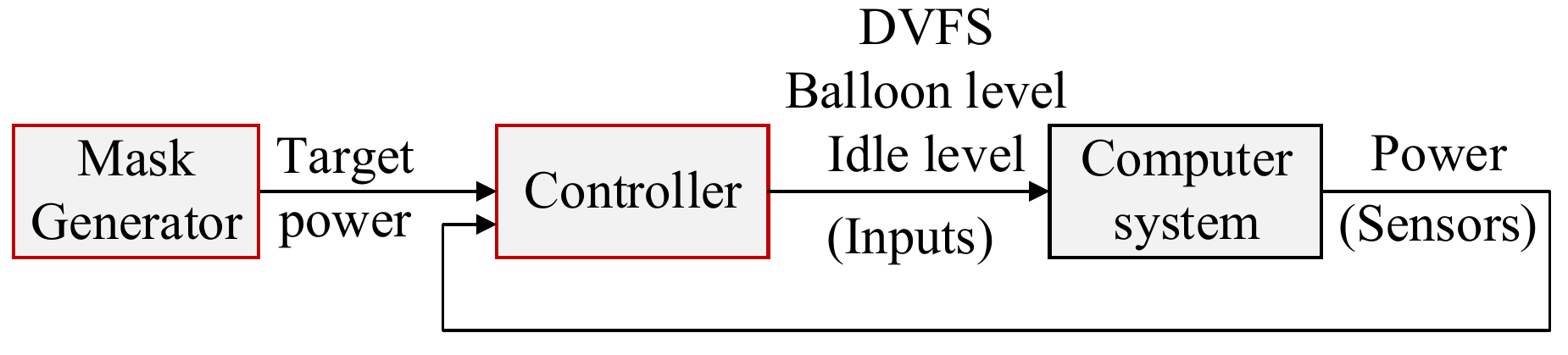}
\caption{High-level architecture of a system that falsifies the power trace
of an application.}
\label{fig_arch}
\vspace{-2mm}
\end{figure}

The high-level architecture of a system that uses control theory techniques
to reshape the power trace of a computer system is shown in Figure~\ref{fig_arch}.
It is composed of a {\em Mask Generator} and a {\em Controller}.
The Mask Generator decides what should  the target power be at each time, so that
it can mislead any attacker. It continuously communicates 
its target power to the controller. 
In the example of Section~\ref{why}, the Mask Generator would pass the constant
value $P$ to the controller. 
Section~\ref{gen_mask} discusses more advanced cases
that involve passing a time-varying function.

The controller reads this target and the actual power consumed by the computer
system, as given by power sensors. Then, based on its current state, it actuates
on various 
inputs of the computer system, which will bring the power close to the
target power. Some of the possible actuations are: changing the frequency and voltage
of the computer system, and scheduling the balloon thread or the idle thread.



The space of power side-channel attack environments is broad,
which calls for different architectures. Table~\ref{tab_choices}
shows two representative environments, which we call {\em Conventional} and
{\em Specialized} environments. The Conventional environment
is one where the attacker extracts information
through off-the-shelf sensors such as power counters or OS API~\cite{mobilePowerside}.
Such sensors are typically reliable only  at 
coarse measurement granularities -- e.g., every 20ms. Hence, we can use a typical
matrix-based controller, like the one described in Section~\ref{sub_robust},
which can respond in 5--10$\mu$s. Given these timescales, the 
controller can actuate on parameters such as frequency or 
voltage, or schedule balloon or idle threads. Such controllers can be used
to hide what application is running, or what keystroke is being 
pressed.

\begin{table}[h]
\caption{Two types of power side-channel environments.}
\label{tab_choices}
\footnotesize
\centering
\begin{tabulary}{\columnwidth}{@{}p{0.25\columnwidth}p{0.35\columnwidth}p{0.35\columnwidth}@{}}
\toprule
Characteristic & Conventional   & Specialized  \\
\toprule
Attacker's sensing & Reads power sensors like counters &  Uses devices like oscilloscopes\\
\midrule
Sensing rate & $\approx$ 20ms &  < 50ns\\
\midrule
Controller type & Matrix-based controller, in hardware or privileged software 
& Table-based controller in hardware \\
\midrule
Controller Response Time &  5--10$\mu$s & $\approx$ 10ns \\
\midrule
Example Actuations  & Change frequency and voltage; schedule balloon and idle threads & 
Insert instructions and pipeline bubbles \\
\midrule
Example Use Cases & Hide what application runs or what keystroke occurred &  
Hide features of a crypto algorithm\\
 \bottomrule
\end{tabulary}
\end{table}

The Specialized environment is one where the attacker extracts information
using specialized hardware devices, such as oscilloscopes. 
The frequency of samples can be in the tens of nanoseconds. In this case, the
controller has to be very fast. Hence, it cannot use the matrix-based
approach of Section~\ref{sub_robust}. Instead, tt has to rely on a table of 
pre-computed values. Its operation
involves a table look-up that determines what action to take. 
This controller is implemented in hardware and has a response
time of no more than around 10ns. 

A possible design involves actuating
on a hardware module that immediately inserts compute instructions into the ROB or pipeline bubbles to mislead any attacker. With such a
fast actuation, this type of defense can be used, for example,
to hide the features of a cryptographic algorithm.

In the rest of the paper, we focus on the first type of environment only,
as it is by far the easiest to mount and widely used.



\begin{figure*}[h]
\centering
\subfloat{
\includegraphics[width=0.19\textwidth]{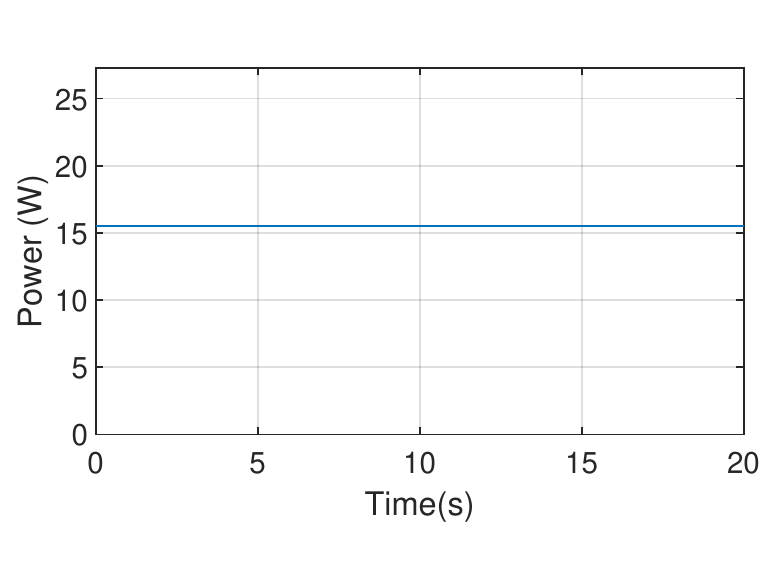}
\label{subfig_c}
}
\subfloat{
\includegraphics[width=0.19\textwidth]{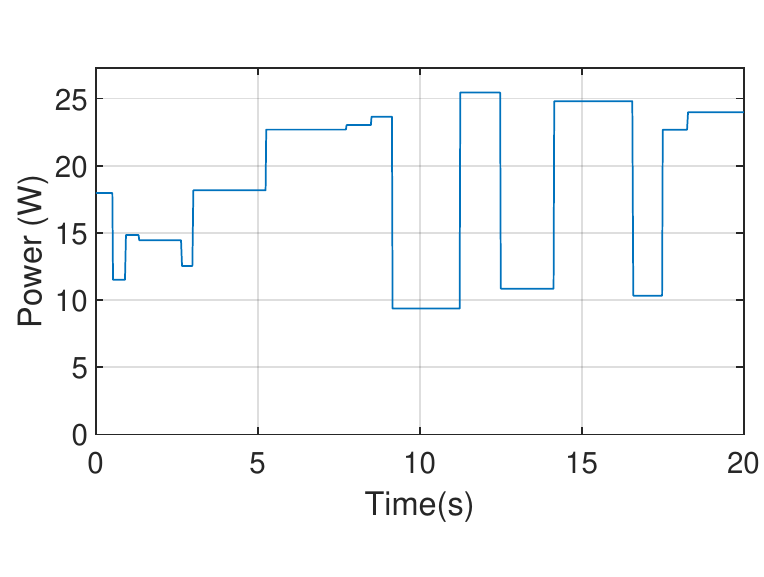}
\label{subfig_u}
}
\subfloat{
\includegraphics[width=0.19\textwidth]{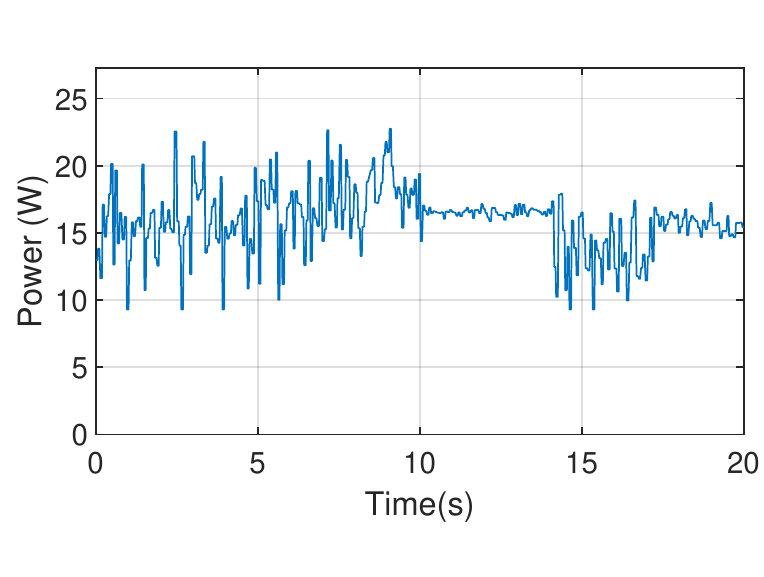}
\label{subfig_n}
}
\subfloat{
\includegraphics[width=0.19\textwidth]{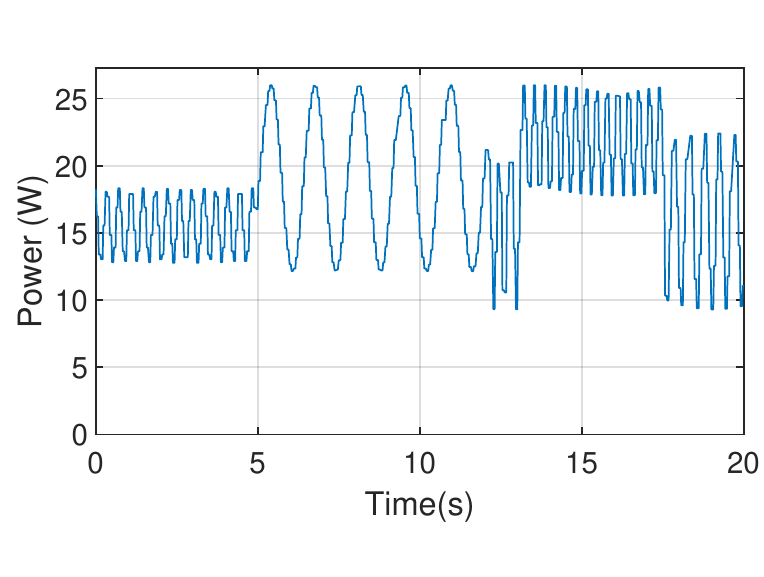}
\label{subfig_s}
}
\subfloat{
\includegraphics[width=0.19\textwidth]{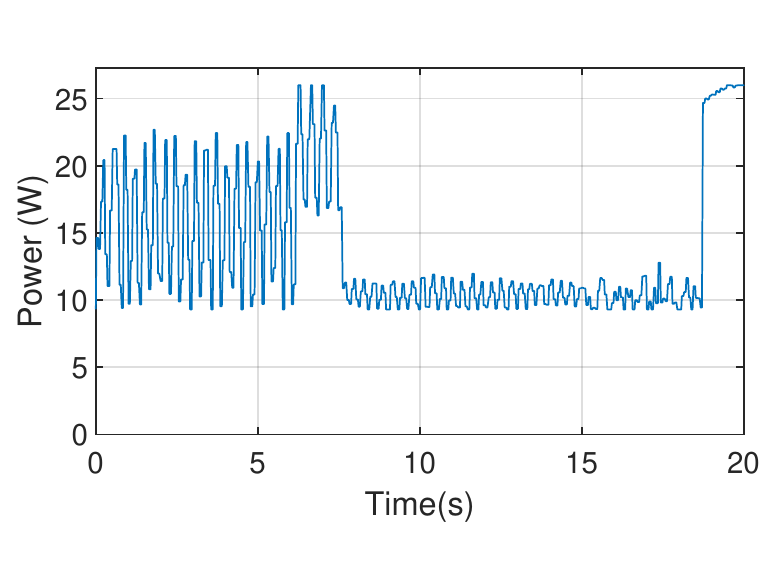}
\label{subfig_ns}
}\vspace{-5mm}\\
\setcounter{subfigure}{0}
\subfloat[Constant]{
\includegraphics[width=0.19\textwidth]{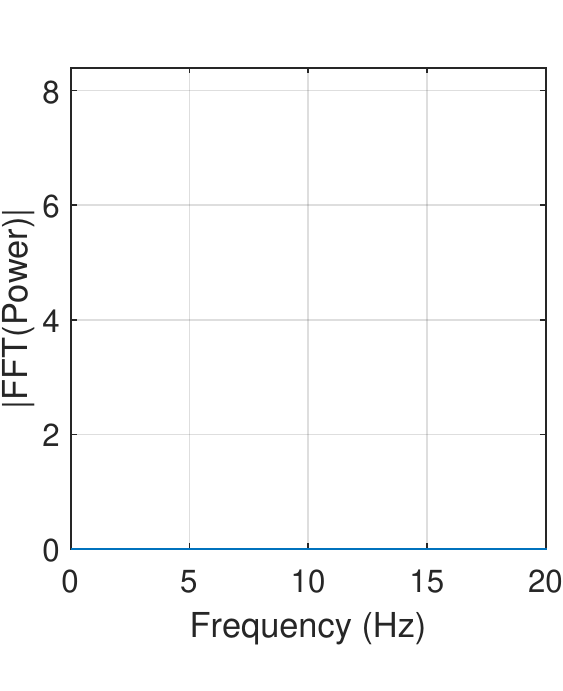}
\label{subfig_cf}
}
\subfloat[Uniformly Random]{
\includegraphics[width=0.19\textwidth]{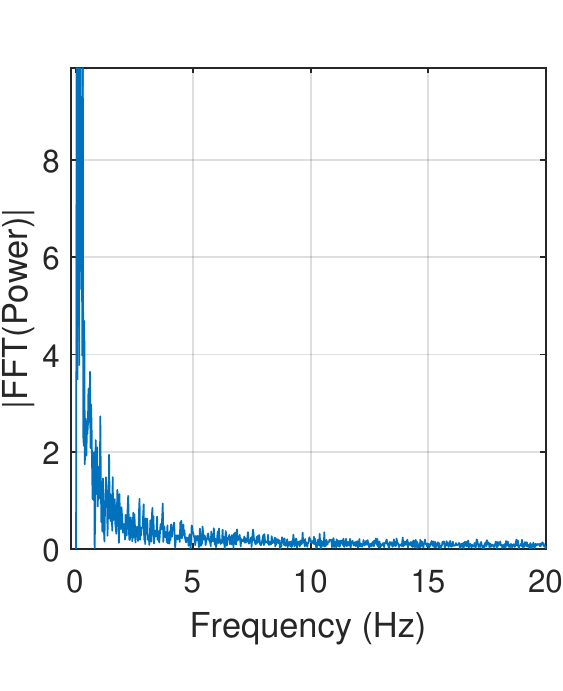}
\label{subfig_uf}
}
\subfloat[Gaussian]{
\includegraphics[width=0.19\textwidth]{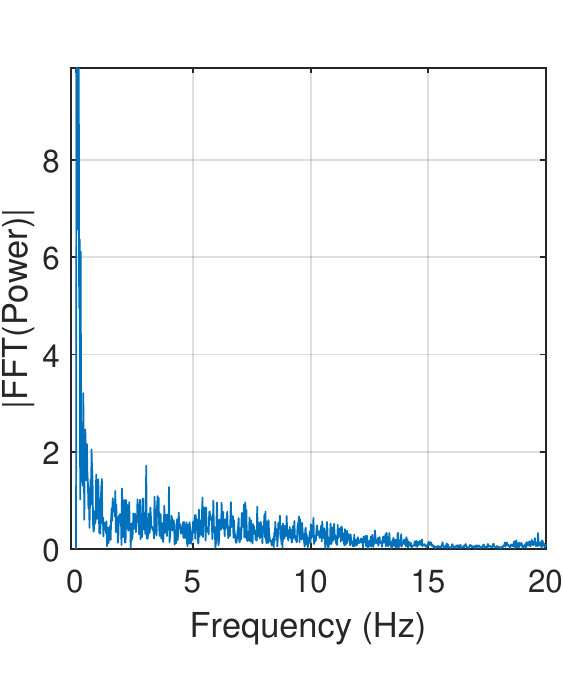}
\label{subfig_nf}
}
\subfloat[Sinusoid]{
\includegraphics[width=0.19\textwidth]{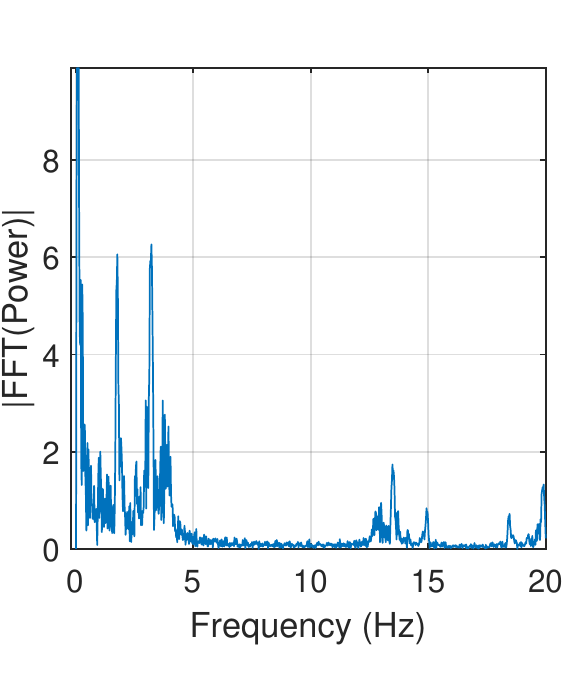}
\label{subfig_sf}
}
\subfloat[Gaussian Sinusoid]{
\includegraphics[width=0.19\textwidth]{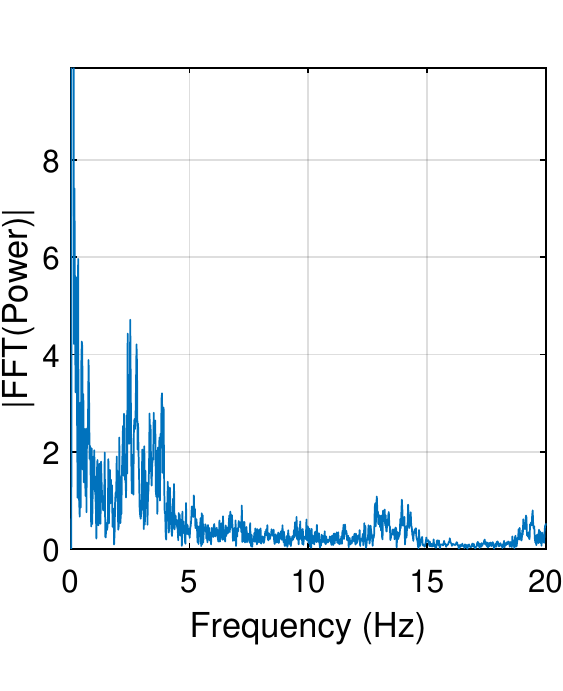}
\label{subfig_nsf}
}
\caption{Examples of different masks. In each case, the time-domain curve is
at the top, and the frequency-domain one at the bottom.\label{masks}}
\vspace{-5mm}
\end{figure*}

\subsection{Generating Effective Masks}
\label{gen_mask}


To effectively mislead an attacker,
it is not enough for the defense to only be able to track power
targets closely (as
discussed in Section~\ref{why}). In addition, the defense must create an appropriate target power signal (i.e., an
appropriate {\em mask}). The module that determines the mask to be used at each interval is the Mask Generator.


An effective mask must protect information leaked in both time domain and the frequency domain (i.e., after obtaining its FFT) because attackers can analyze signals in either domain.

We postulate that, to be effective, 
a mask must have three properties. First, its mean and variance 
must change over the time domain, to portray phase behavior
(Figure~\ref{masks}(c) top). 
Such changes will mask the original signal in the time domain.


The second property is that the mean and variance changes must have 
various rates -- from smooth to abrupt. This property will cause the
resulting frequency domain curve to {\em spread} over a range of 
frequencies (Figure~\ref{masks}(c) bottom). As a result, the curve will a property
similar to typical curves generated natively by applications.

The final property is that the target signal must have repeating 
patterns at various rates. This is to create various {\em peaks} in the 
frequency domain curve (Figure~\ref{masks}(e) bottom). Such peaks are common
in applications, as they represent the effects of loops.

Table~\ref{tab_signals} lists some well-known signals, showing
whether each signal
changes the mean and the variance in the time domain, and whether it
creates spread and peaks in the frequency domain.
Such properties determine their viability 
to be used as effective masks. Figure~\ref{masks} 
shows a graphical representation of each signal in order.

\begin{table}[h]
\caption{Some standard signals and what they change in the time and frequency domains.}
\label{tab_signals}
\footnotesize
\centering
\begin{tabulary}{\columnwidth}{@{}p{0.35\columnwidth}p{0.08\columnwidth}p{0.11\columnwidth}p{0.14\columnwidth}p{0.14\columnwidth}@{}}
\toprule
& \multicolumn{2}{c}{Time-domain} & \multicolumn{2}{c}{Frequency-domain} \\
\cmidrule(r){2-3} \cmidrule(l){4-5}
Signal & Mean &  Variance & Spread  & Peaks \\
\toprule
Constant & -- & -- & -- & -- \\
Uniformly Random & Yes & -- & Yes & -- \\
Gaussian  & Yes & Yes  & Yes & -- \\
Sinusoid & Yes & Yes & --  & Yes\\
Gaussian Sinusoid & Yes & Yes & Yes & Yes\\
 \bottomrule
\end{tabulary}
\end{table}

A {\em Constant} signal (Figure~\ref{masks}(a)) 
does not change the mean or variance in the time domain, or
create  spread or peaks in the frequency domain.
Note that this signal cannot be realized in practice. 
Any realistic method of keeping the output signal constant
under changing conditions would have to first observe the outputs deviating from the targets, and then set the inputs accordingly. Hence, the output signal would
have a burst of power activity at all the change points in the application. 
As a result, such signal would easily leak information.
Furthermore, the signals obtained in multiple runs of a given 
application would be similar to each other.


In a {\em Uniformly Random} signal (Figure~\ref{masks}(b)),
a value is chosen randomly from a range, and is used as a target for a random duration
in the time domain. After this period, another value and duration are selected, and the process repeats. 
This signal changes the mean but not the variance in the time domain. 
In the frequency domain, the signal is spread across a range but has no peaks. 
This mask is not a good choice either 
because any repeating activity in the application would be hard to hide
in the time domain signal. 

The {\em Gaussian} signal (Figure~\ref{masks}(c))
takes a gaussian distribution and 
keeps changing the mean and variance randomly in the
time domain. The resulting frequency-domain signal is spread over
multiple frequencies, but does not have peaks.


The {\em Sinusoid} signal (Figure~\ref{masks}(d))
generates a sinusoid and keeps changing the
frequency, the amplitude, and the offset (i.e., the 
power at angle 0 of the sinusoid)
randomly with time. This signal changes the
mean and variance in the time domain. In the frequency
domain, it has clear sharp peaks at each of its sine wave frequencies.
However, there is no spread. Therefore, this signal is not effective at 
masking abrupt changes in the native power output of applications. 


Finally, the {\em Gaussian Sinusoid} (Figure~\ref{masks}(e))
is the addition of the previous two signals. This signal has all the
properties that we want. The mean and variance in the time domain, and the peak
locations in the frequency domain are varied by changing the parameters of the sinusoid. The gaussian component widens the peaks in the frequency domain, causing spread.
This is the mask that we propose.

\section{Implementation on Two Systems}
\label{implement}

We implement {\em \des} as the system shown in Figure~\ref{fig_arch}.
We target the {\em Conventional} environment, shown in the center of Table~\ref{tab_choices},
as it is by far the most frequent one. We implement
\des in software, as OS threads, in two different machines. 
System One ({\em Sys1}) is a consumer class machine with 6 cores. 
Each core supports 2 hardware contexts, 
totaling 12 logical cores. 
System Two ({\em Sys2}) is a server class machine with 2 sockets, 
each having 10 cores of 2 hardware contexts, for a total of 40 logical cores.


On both systems, the processors are  Intel Sandybridge. 
The OS is CentOS 7.6, based on the Linux kernel version 3.10.  
On each context, we run one of three possible threads: a thread of
a parallel benchmark, an idle thread, or a balloon thread. The 
latter is one thread of a parallel program that we call the {\em Balloon}
program, which performs floating-point array operations in a loop
to raise the power consumed. 
Both the benchmark and the Balloon program have
as many threads as logical cores.

In each system, the controller measures one output and actuates on
three inputs. The output is the total 
power consumed by the chip(s). The inputs are the
DVFS level applied, the percentage of idle thread execution, and the
percentage of balloon program execution.

DVFS values can be changed from 1.2 GHz to 2.0 GHz on Sys1,
and from 1.2 GHz to 2.6 GHz on Sys2, with 0.1 GHz increments in either case. 
On both systems, idle thread execution can be set from 0\% to 48\% 
in steps of 4\%, and the balloon program execution from 0\% to 100\% in steps of 
10\%.

Power consumption is measured through RAPL interfaces~\cite{intelRapl}.
The DVFS level is set using the \texttt{cpufreq} interface. The 
idle thread setting is specified through Intel's Power Clamp interface~\cite{powerclamp}. 
The balloon program setting is specified through an \texttt{shm} file.
The controller, mask generator, and balloon program run as privileged 
processes.

\subsection{Designing the Controller}
\label{sub_robust_design}

We design the controller using robust control theory~\cite{mimoText}.
To develop it, we need to: (i) design a dynamic model of the computer 
system running the
applications, and (ii) set three parameters of the controller 
(Section~\ref{sub_robust}), namely
the input weights, the uncertainty guardband, and
the output deviation bounds~\cite{mimoText}. 

To develop the model, we use the System Identification~\cite{sysidText}
experimental modeling methodology. In this approach, we run training 
applications on the computer system and, during execution, change 
the system inputs. We log the observed outputs and the inputs. 
From the data, we construct a dynamic polynomial model of the computer
system:

\begin{equation}
\begin{split}
y(T) = a_1\times y(T-1) + \ldots + a_m\times y(T-m) + \\
b_1\times u(T) + \ldots + b_n\times u(T-n+1) 
\end{split}
\label{eq_model}
 \end{equation}
In this equation, y(T) and u(T) are the outputs and inputs, respectively, at time T. This model describes the outputs at any time T as a function of the {\em m} past outputs, 
and the current and {\em n-1} past inputs. The constants $a_i$ and $b_i$
are obtained by least squares minimization from the experimental data~\cite{sysidText}.

We perform system identification by running 
two applications from PARSEC 3.0 (\textit{swaptions} and \textit{ferret})
and two from SPLASH2x (\textit{barnes} and \textit{raytrace})~\cite{parsec}
on Sys1. The models we obtain have a dimension of 4 (i.e., $m=n= 4$ in Equation~\ref{eq_model}). The system identification approach is powerful to capture the relationship between the inputs and outputs.

The input weights are set depending on the relative overhead of 
changing each input. In our system, all inputs have similar changing
overheads. Hence, we set all the input weights to 1. 
Next, we specify the uncertainty guardband by evaluating several choices. For each uncertainty guardband
choice, Matlab tools~\cite{robustMATLAB} give the smallest output deviation bounds the controller can provide. Based on prior work~\cite{mimo16isca},
we set the guardband to be 40\%, which allows the  
output deviation bounds for power to be within 10\%. 

With the model and these specifications, standard tools~\cite{robustMATLAB} generate the
set of matrices that encode the  controller (Section~\ref{sub_robust}). This controller's dimension is 11 i.e. the number of elements in the state vector of Equation~\ref{eq:control} is 11. The controller runs periodically every 20 ms. We set this duration based on the 
update rate of the power sensors.




\subsection{Mask Generator}
\label{sub_mask_design}

As indicated in Section~\ref{gen_mask}, our choice of mask  is a gaussian
sinusoid (Figure~\ref{masks}(e)). This signal is the sum of a sinusoid and a gaussian,
and its value as a function to time (T) is: 
\begin{equation}
\begin{split}
Offset + Amp \times \sin(\frac{2\pi \times T}{Freq}) + Noise(\mu, \sigma)
\end{split}
\label{eq_ns}
\end{equation}
where the Offset, Amp, Freq, $\mu$ and $\sigma$ parameters keep changing. Each of these parameters is selected at random from a range of values, subject to two constraints.
First, the maximum power cannot be over the Thermal Design Power (TDP) of the system.
Second, the frequency has to be smaller than half of the power sampling frequency;
otherwise, the sampler would not be able to identify the curve.
Once a particular set of parameters is chosen, the mask generator  uses them for 
$N_{hold}$ samples, after which the parameters are updated again. $N_{hold}$ 
itself varies randomly between 6 to 120 samples. 

\section{Evaluation Methodology}
\label{setup}


We analyze the security offered by \des in two ways. First, we consider two machine learning-based power analysis attacks and evaluate how \des can prevent them. Second, we use signal processing metrics to evaluate the power-shaping properties of \des. 

\subsection{Machine Learning Based Power Attacks}
\label{sub_attacks}

Pattern recognition is at the core of nearly all power analysis attacks~\cite{mobilePowerside,vmEnergyAttack,batteryAttack,iotDeepLearnEm,Kocher2011}. Therefore, we use multiple machine learning-based attacks to test \des.

\noindent
\textbf{1. Detecting the Active Application:} This is a fundamental attack that is reported in several works~\cite{mobilePowerside,vmEnergyAttack,batteryAttack,iotDeepLearnEm}. The goal is to infer the application running on the system using power traces. Initially, attackers gather several power traces of the applications that must be detected, and train a machine learning classifier to predict the application from the power trace. Then, they use this classifier to predict which application is creating a new signal from the system. This attack tells the attackers whether a power signal is of use to the them, and enables them to identify more serious information from the power trace. For example, if the attackers know that the power signal belongs to a video encoder, they can infer that the repeating patterns in it correspond to successive frames being encoded.

We implement this attack on \textit{Sys1} using applications from 
PARSEC 3.0 (blackscholes, bodytrack, freqmine, raytrace, and vips)
and SPLASH2x (radiosity, water\_nsquared, and water\_spatial). 
We run each application 1,000 times with native datasets on \textit{Sys1}  and collect a total of 8,000 power traces. We collect power measurements using unprivileged RAPL counters -- an ideal case for an attacker because this does not need physical access and has accurate measurements. For robust classification under noise, we configure the machine in each run with a {\em different frequency and level of idle activity} before launching the application. The idle activity results in a noisy power trace because of the interference between the idle threads and the actual applications.

\begin{figure*}[t]
\centering
\subfloat[Noisy Baseline $\to$ \newline \des Constant (12\%)]{
\includegraphics[width=0.25\textwidth]{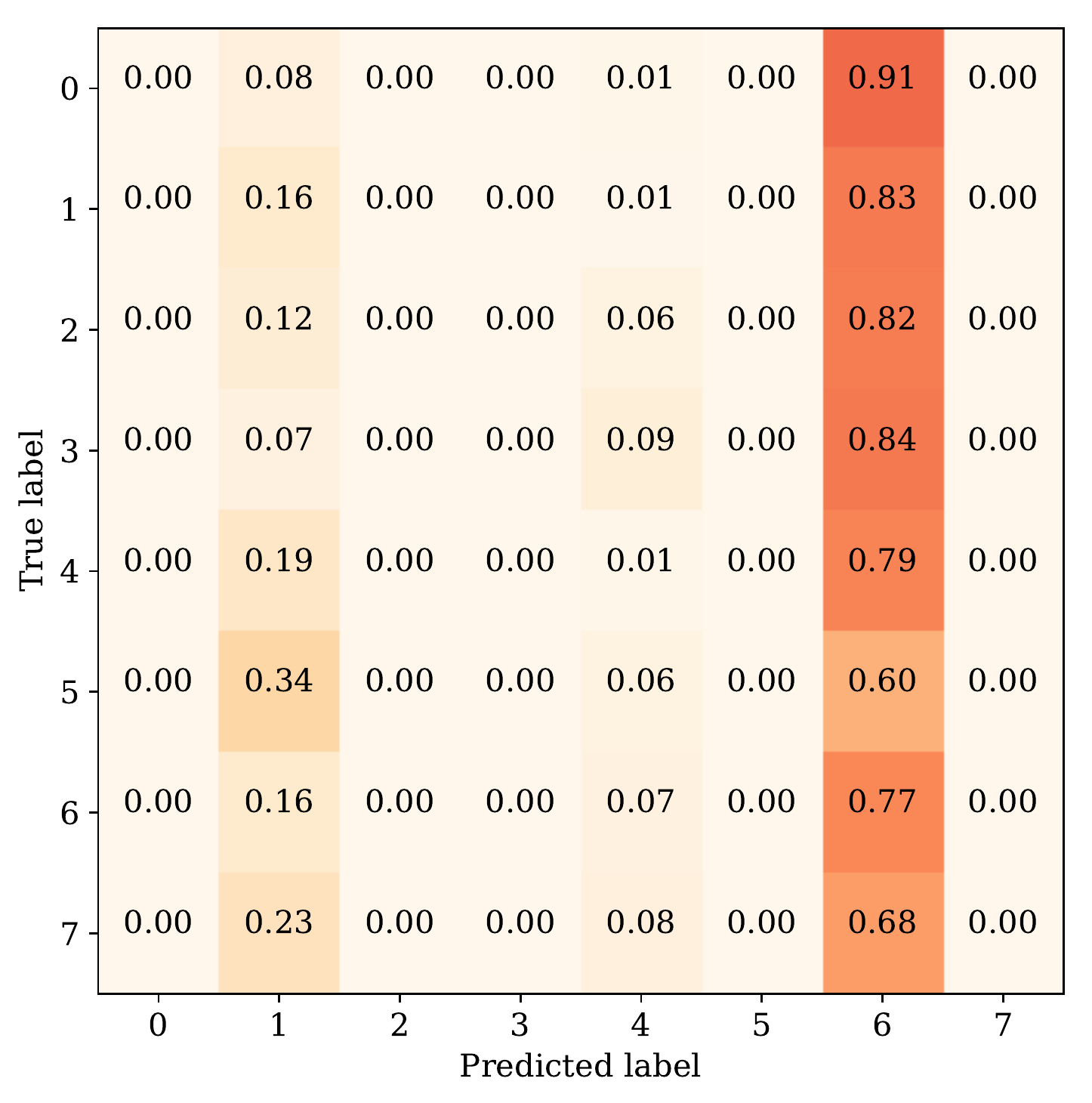}
\label{subfig_attack1pbpc}
}
\subfloat[\des Constant $\to$ \newline \des Constant (67\%)]{
\includegraphics[width=0.25\textwidth]{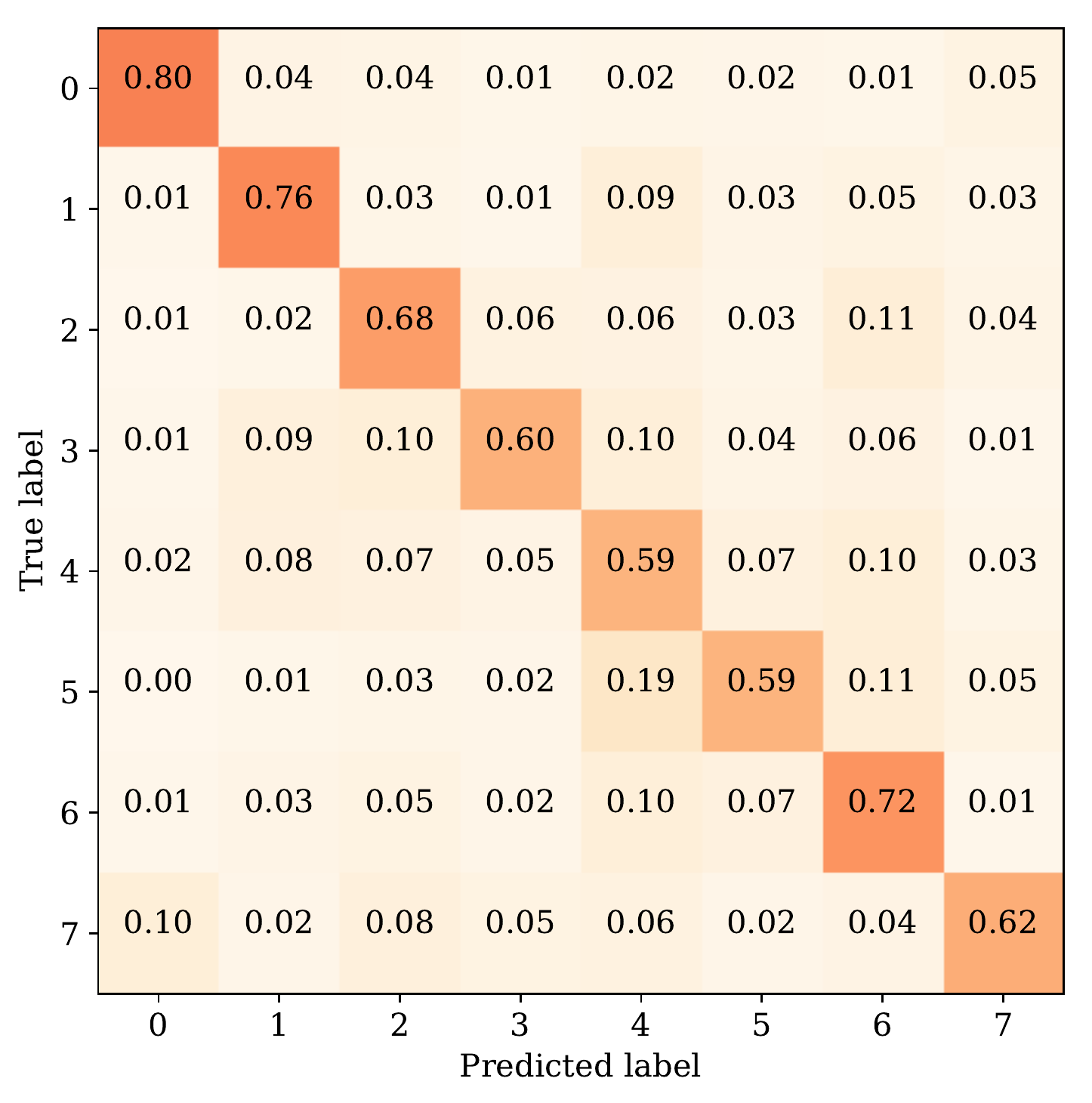}
\label{subfig_adattack1pcpc}
}
\subfloat[Noisy Baseline $\to$ \newline \des Gaussian Sinusoid (13\%)]{
\includegraphics[width=0.25\textwidth]{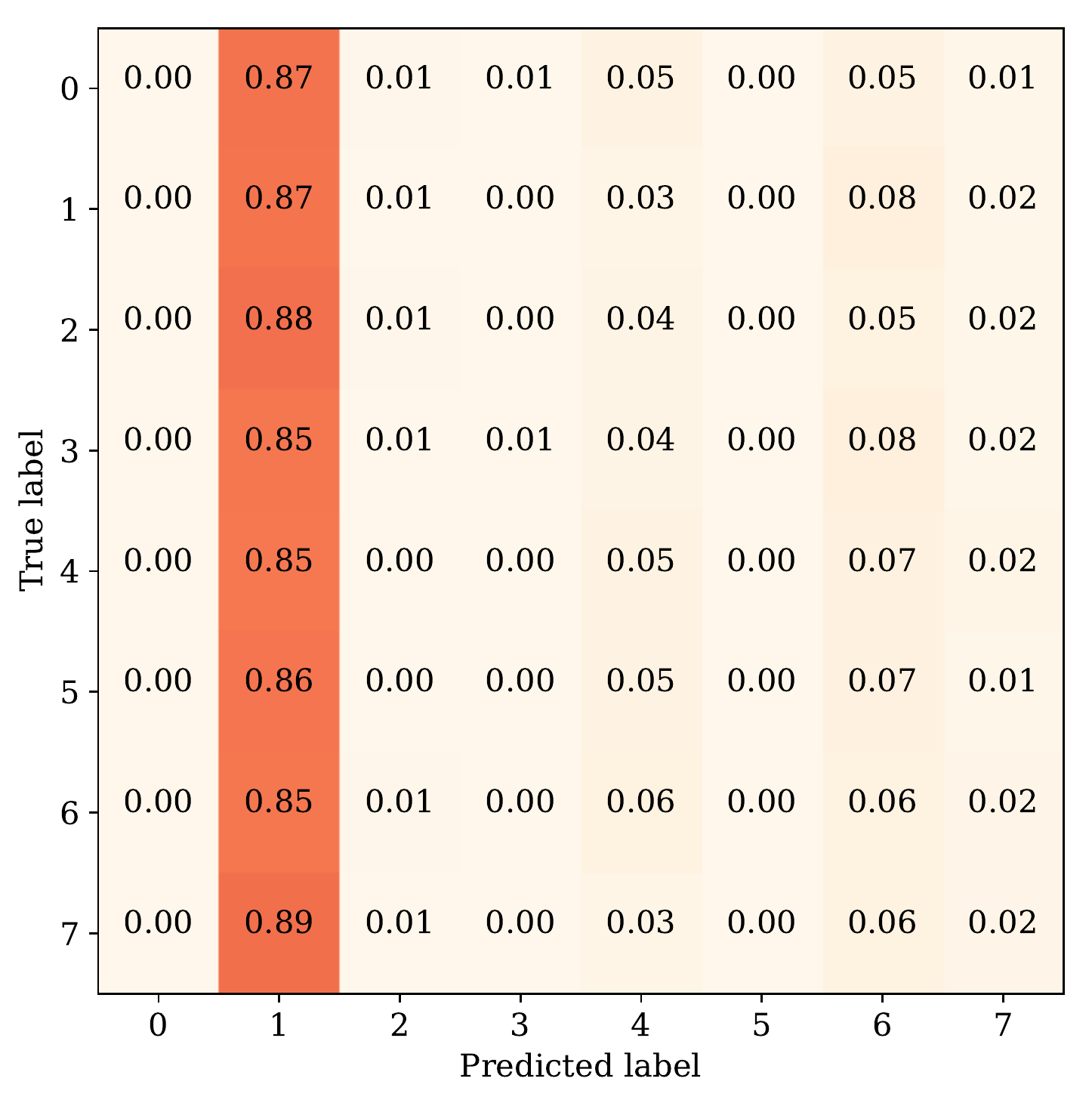}
\label{subfig_attack1pbpns}
}
\subfloat[\des Gaussian Sinusoid $\to$ \des Gaussian Sinusoid (20\%)]{
\includegraphics[width=0.25\textwidth]{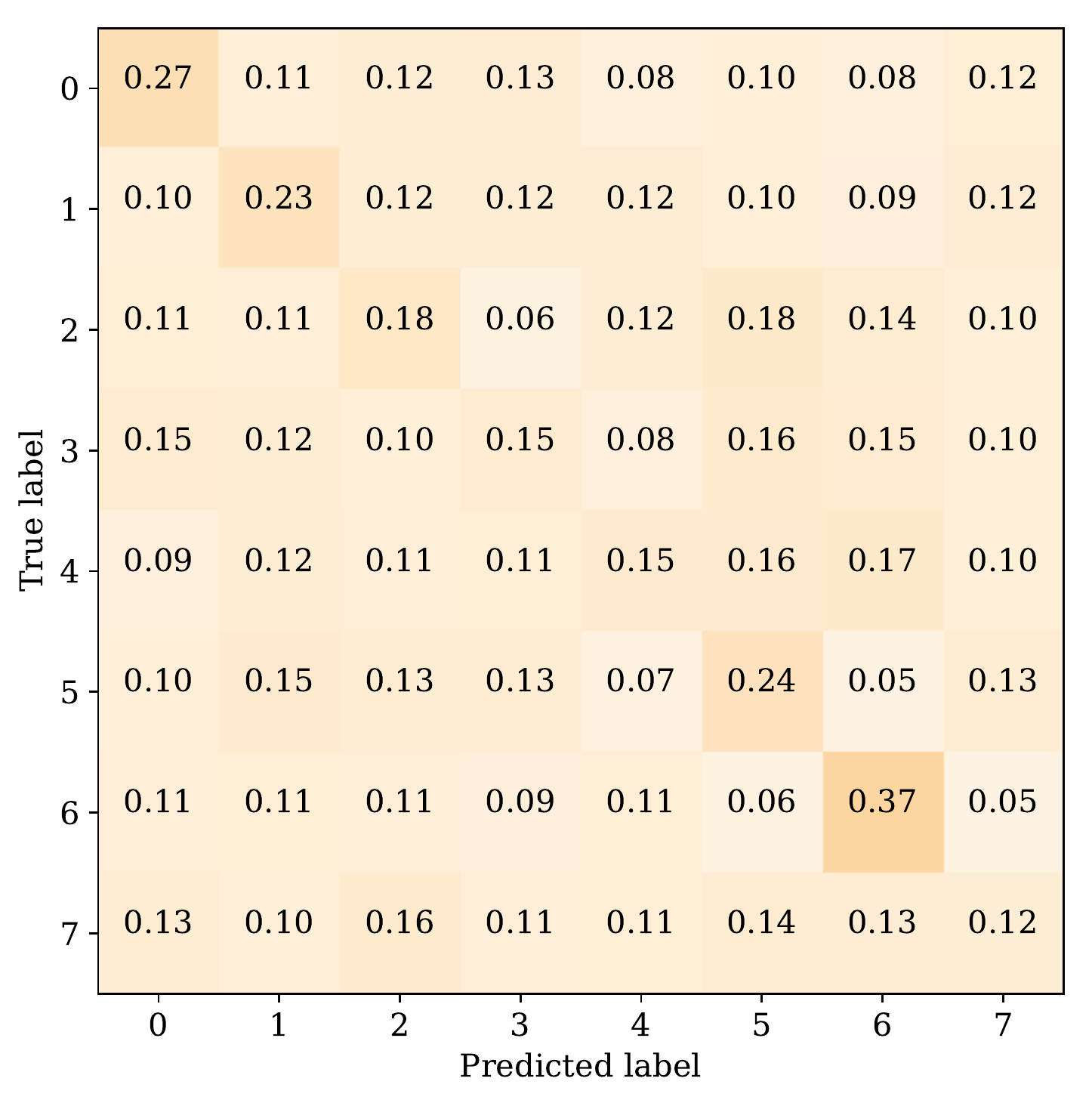}
\label{subfig_adattack1pnspns}
}
\caption{Confusion matrices for a machine learning attack to identify the active application from power signals. The figures are labeled in the format: Train dataset $\to$ Test dataset and the average accuracy is in parenthesis. Higher fractions are in darker squares.\label{fig_attack1}}
\vspace{-4mm}
\end{figure*}

From each trace, we extract multiple segments of 15,000 RAPL measurements, and average the 5 consecutive measurements in each segment to remove effects of noise. This reduces each segment length to 3,000. Then, we convert the values in the segment into one-hot representation by quantizing the values into one of 10 levels  encoded in one-hot format. This gives us 30,000-long samples that we feed into a classifier. Among the samples from all traces, we use 60\% of them for training, 20\% for validation, and leave 20\% as the test set.

Our classifier is a multilayer perceptron neural network with two hidden ReLU layers of dimensions 1,500 and 600. The output layer uses Logsoftmax and has 8 nodes, corresponding to the 8 applications we classify. With this model,
we achieve a training accuracy of 99\%, and validation and test accuracies of 92\%.

\noindent
\textbf{2. Detecting the Application Data:} This attack is used to infer the differences in data that a given application uses -- such as the websites accessed by a browser, or the videos processed  by an encoder. This is also a common attack described in multiple works~\cite{mobilePowerside,batteryAttack,usbPowerSide,powerspy,iotDeepLearnEm}. We implement this attack on \textit{Sys2} targeting the ffmpeg video encoding application~\cite{ffmpeg}. We use three videos saved in raw format: tractor, riverbed and wind. They are commonly used for video testing~\cite{videolink}. We transcode each video with x264 compression using ffmpeg and record the power trace. Since the power consumed by video encoding depends on the content of the frames, each video has a distinct power pattern. 

We collect the power traces from 300 runs of encoding each video with different frequency and idle activity levels. Next, we choose multiple windows of 1,000 measurements from each trace. As with the previous attack, we average 5 consecutive measurements and use on-hot encoding to obtain samples that are 2,000 values long.  

The classifier for this attack is also a multilayer perceptron neural network with two hidden ReLU layers of dimensions 100 and 40. The output layer uses Logsoftmax and has 3 nodes corresponding to the 3 videos we classify. 
With this model, we achieve training, validation and test accuracies of 99\%.

\noindent
\textbf{3. Adaptive Attacks:} We consider an advanced scenario where the attacker records the distorted power traces of applications when \des is running, and knows
which application is running. She then trains models to perform the previous attacks. The data collection and model training is the same as described above.



\subsection{Signal Processing Analysis}
\label{sub_metrics}

In addition to the machine learning attacks, we analyze \des's distortion using 
the following signal processing metrics.

\noindent
\textbf{1. Signal Averaging:} Averaging multiple power signals removes random noise in the signals, allowing attackers to detect even small changes. We test this using signal averaging analysis on \textit{Sys1}. For each application, we collect three sets of
1,000 noisy power signals with: (i) no \des, (ii) \des  Constant, and (iii) \des Gaussian Sinusoid. When \des is not used, noise is created by changing frequency and idle activity. Then, we average the signals for each application and analyze the distribution of values in the averaged signals. Effective obfuscation would cause the values in the averaged traces to be distributed in the same manner across applications.

\noindent

\noindent
\textbf{2. Change Detection:} This is a signal processing technique used to identify times when the properties of a signal change. The properties can be the signal mean, variance, edges, or fourier coefficients. We use a standard Change Point Detection algorithm~\cite{changepts} to compare the change points found in the baseline and in the re-shaped signals on \textit{Sys2}.

\section{Evaluating \des}
\label{results}


\subsection{Evading Machine Learning Attacks}
\label{sub_part1}

\noindent
\textbf{Application Detection Attack:} Figure~\ref{fig_attack1} shows the confusion matrices for performing the application detection attack and its adaptive variant on \textit{Sys1}. The rows of the matrix correspond to the true labels of the traces and the columns are the predicted labels by the machine learning models. The 8 applications we detect are numbered from 0 to 7. A cell in the i$^{th}$ row and j$^{th}$ column of the matrix lists the fraction of traces belonging to label i that were classified as j. Values close to 1 along the diagonal indicate accurate prediction. We use {\em Baseline} to refer to the environment without \des, and {\em Obfuscated} to
refer when
\des runs.

Recall that the model trained on Noisy Baseline signals can classify unobfuscated signals with 92\% accuracy. When this classifier is applied to signals produced when \des runs with a Constant mask, the accuracy drops to  12\% (average of the diagonal values in Figure~\ref{subfig_attack1pbpc}). However, when an advanced attacker trains on the obfuscated traces, the classification accuracy is 67\% (Figure~\ref{subfig_adattack1pcpc}). 

The poor security offered by a Constant mask in the advanced attack is due to two reasons. First, it cannot hide the natural power variations of an application as described in Section~\ref{gen_mask}. Second, the power signals of an application with a Constant mask are similar across multiple runs. Hence, a Constant mask is ineffective at preventing information leakage.

Next, consider the Gaussian Sinusoid mask. When a Noisy Baseline-trained classifier is used, the accuracy is 13\% (Figure~\ref{subfig_attack1pbpns}). Even in an Adaptive attack that trains on the obfuscated traces, the classification accuracy is only 20\% (Figure~\ref{subfig_adattack1pnspns}). This indicates an excellent obfuscation, considering that the random-chance of guessing the correct application is 13\%. The Gaussian Sinusoid introduces several types of variation in the power signals, effectively eliminating any original patterns. Moreover, each run with this mask generates a new form. Therefore, there is no common pattern between different runs that a machine learning module could learn.

\begin{figure}[h]
\centering
\subfloat[Noisy Baseline $\to$ \newline \des Constant (21\%)]{
\includegraphics[width=0.35\columnwidth]{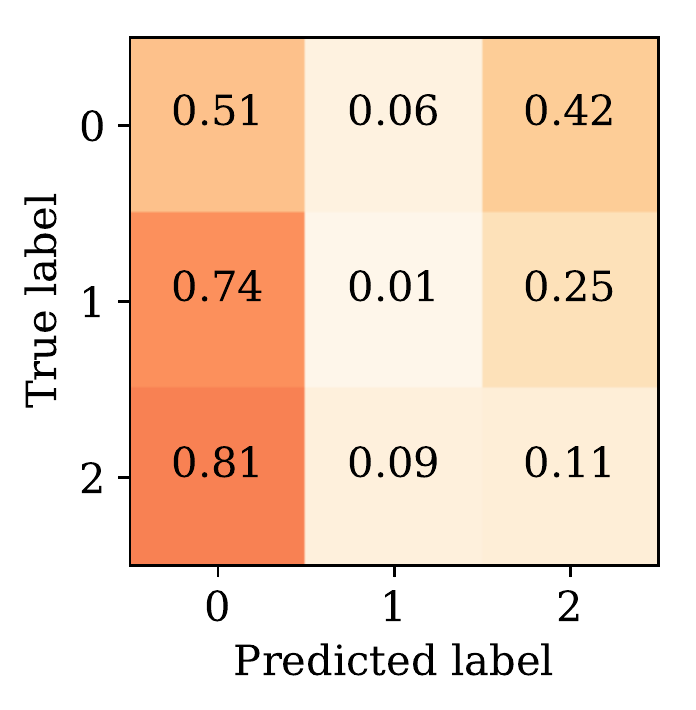}
\label{subfig_attack2vbvc}
}\hspace{1cm}
\subfloat[\des Constant $\to$ \des Constant (79\%)]{
\includegraphics[width=0.35\columnwidth]{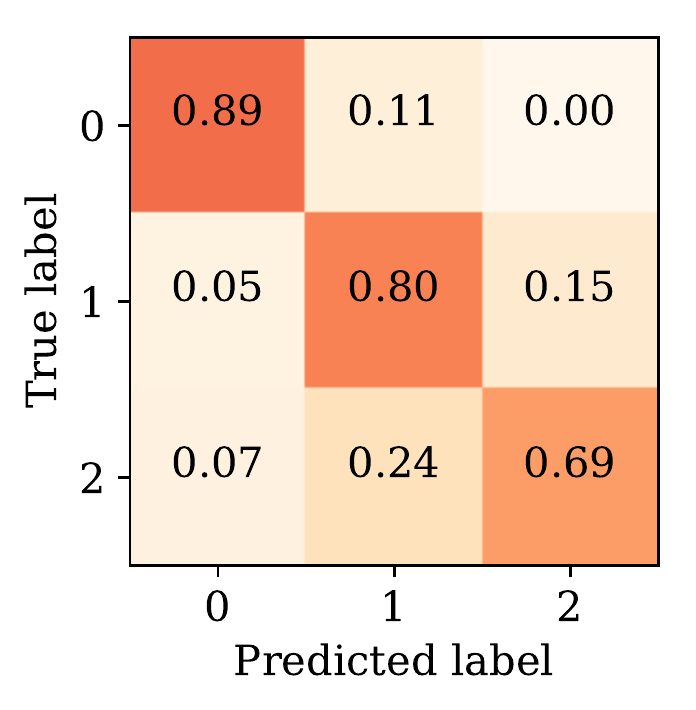}
\label{subfig_adattack2vcvc}
}\\
\subfloat[Noisy Baseline $\to$ \des Gaussian Sinusoid (34\%)]{
\includegraphics[width=0.35\columnwidth]{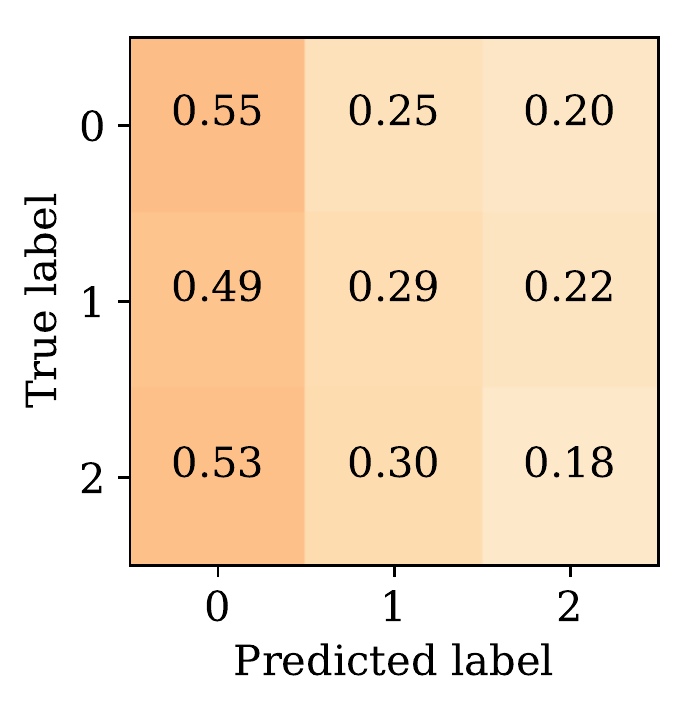}
\label{subfig_attack2vbvns}
}\hspace{1cm}
\subfloat[\des Gaussian Sinusoid $\to$ \des Gaussian Sinusoid (39\%)]{
\includegraphics[width=0.35\columnwidth]{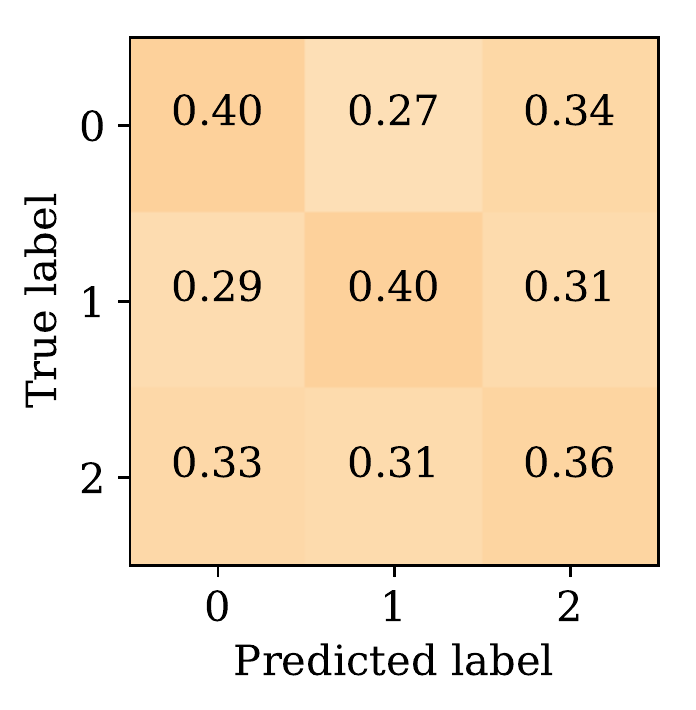}
\label{subfig_adattack2vnsvns}
}
\caption{Confusion matrices for a machine learning attack to identify the video being encoded. The figures are labeled as in Figure~\ref{fig_attack1}. \label{fig_attack2}}
\end{figure}

\noindent
\textbf{Video Data Detection Attack:} Figure~\ref{fig_attack2} shows the confusion matrices for the attack that identifies the video being encoded on \textit{Sys2}. Recall that the attack has to choose among one of three videos for each power trace. The classifier trained on baseline signals had a 99\% test accuracy in predicting the video from the baseline traces. The classification accuracy when using this model on signals with \des's Constant mask is 21\% (Figure~\ref{subfig_attack2vbvc}). However, the accuracy rises to 79\% when the attacker is able to train with the power signals generated by \des using the constant mask (Figure~\ref{subfig_adattack2vcvc}).

\begin{figure*}[htb]
	\centering
	\subfloat[Noisy Baseline]{
		\includegraphics[width=0.3\textwidth]{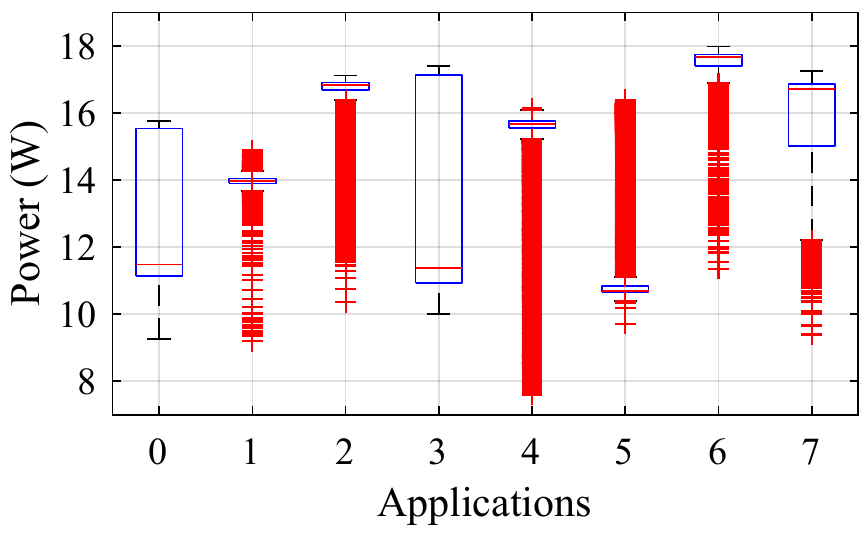}
		\label{subfig_pbbox}
	}
	\subfloat[\des Constant]{
		\includegraphics[width=0.3\textwidth]{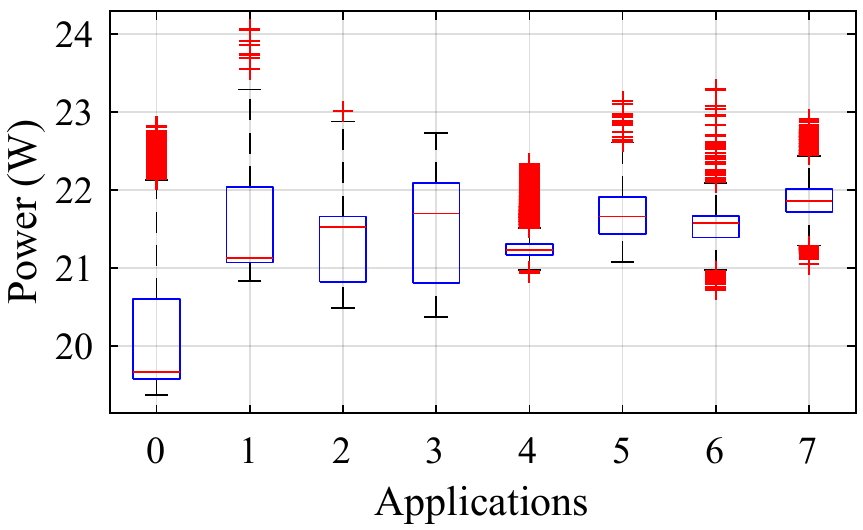}
		\label{subfig_pcbox}
	}
	\subfloat[\des Gaussian Sinusoid]{
		\includegraphics[width=0.3\textwidth]{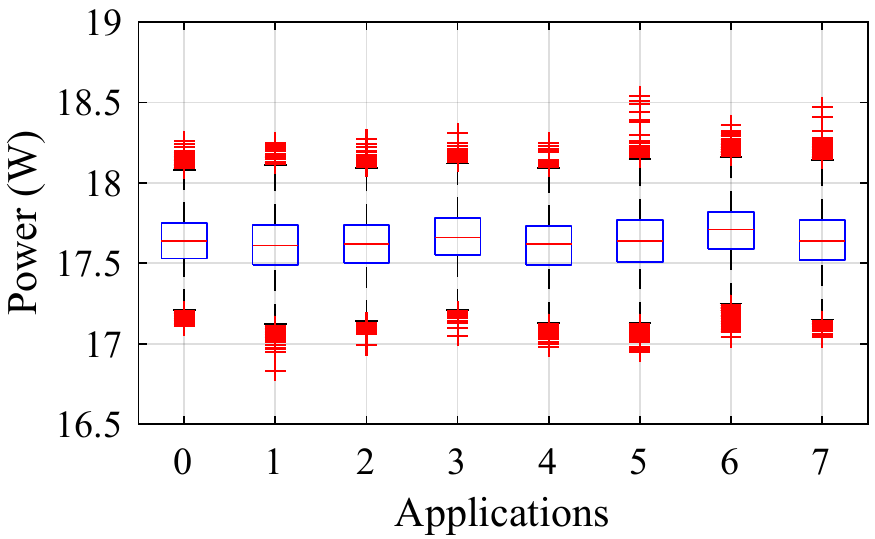}
		\label{subfig_pnsbox}
	}
	\vspace{-2mm}
	\caption{Summary statistics of the average of 1,000 signals. The Y axis of each chart is drawn to a different scale. \label{fig_dist}}
	\vspace{-4mm}
\end{figure*}

\begin{figure*}[htb]
	\centering
	\subfloat[Noisy Baseline]{
		\includegraphics[width=0.3\textwidth]{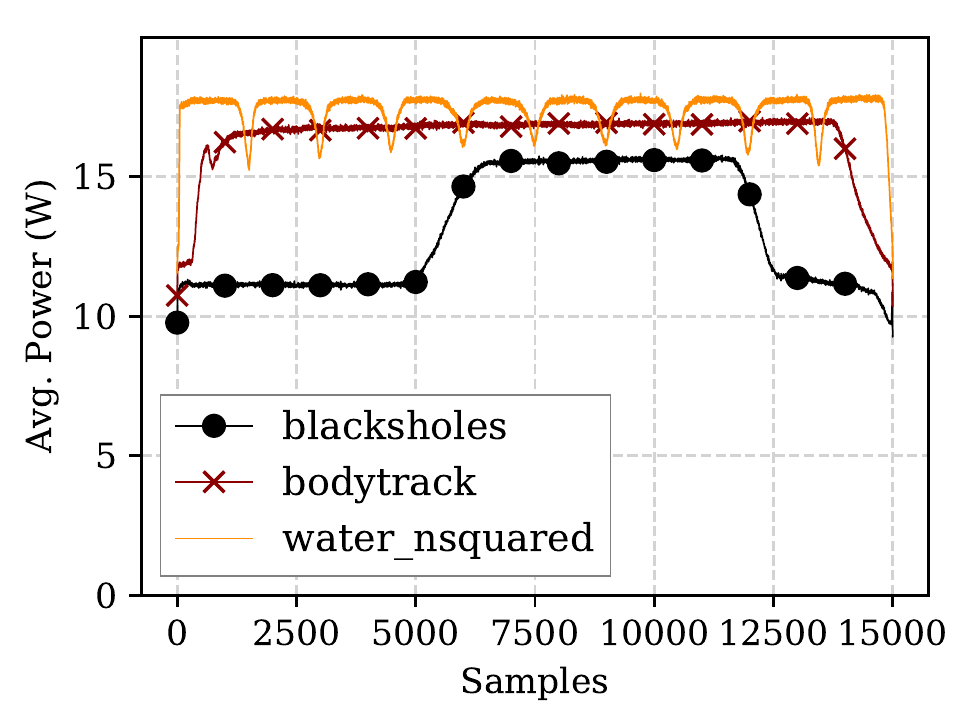}
		\label{subfig_pbav}
	}
	\subfloat[\des Constant]{
		\includegraphics[width=0.3\textwidth]{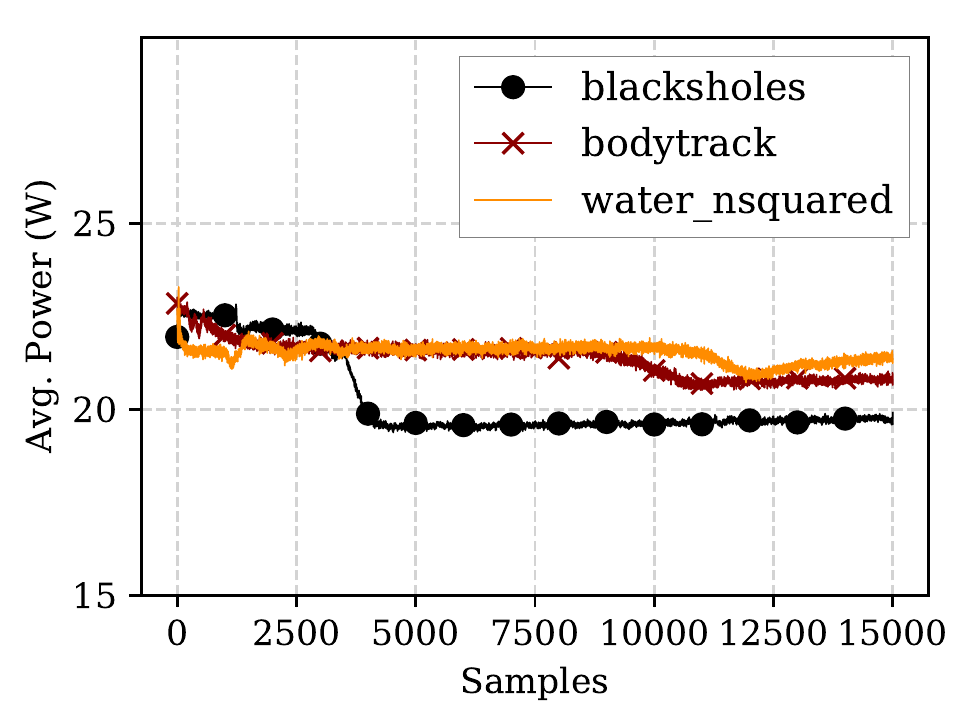}
		\label{subfig_pcav}
	}
	\subfloat[\des Gaussian Sinusoid]{
		\includegraphics[width=0.3\textwidth]{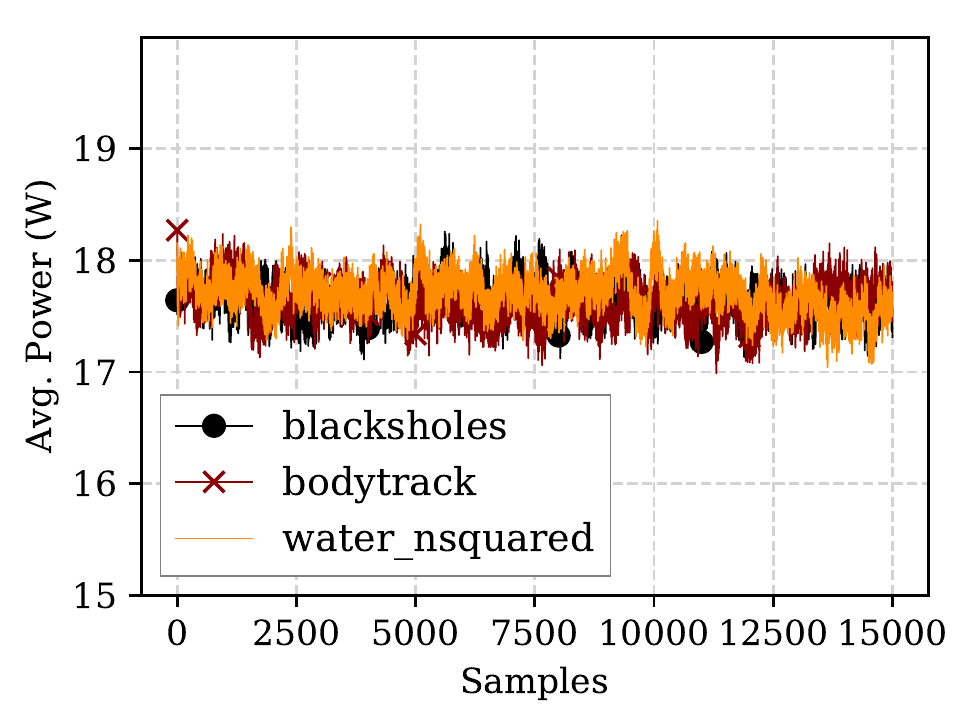}
		\label{subfig_pnsav}
	}
	\caption{Average of 1,000 signals over time samples. The Y axis of each chart is drawn to a different scale. \label{fig_avg}}
	\vspace{-4mm}
\end{figure*}

When the Gaussian Sinusoid mask is used, the classification accuracy when using the Noisy Baseline-trained model is 34\% (Figure~\ref{subfig_attack2vbvns}). Even when the attacker trains with obfuscated traces, the classification accuracy is only 39\%(Figure~\ref{subfig_adattack2vnsvns}). This is a high degree of obfuscation because the random-chance of assigning the correct video to a power signal is 33\%.

Overall, the results from the machine learning-based attacks establish that \des with the Gaussian Sinusoid mask is successful in falsifying pattern recognition attacks. \des's strength is highlighted when it resisted the attacks where the attacker could record thousands of signals generated from \des. This comes from using an effective mask (Gaussian Sinusoid) and a control-theory controller that keeps power close to the mask. Finally, the results also show the Constant mask to be ineffective because it does not have all the needed traits of a mask.

\subsection{Signal Processing Analysis}
\label{sub_part2}

\noindent
\textbf{Signal Averaging:} Figure~\ref{fig_dist} shows the box plots of value distribution in the averaged traces from \textit{Sys1} with the Noisy Baseline, \des Constant mask, and \des Gaussian Sinusoid mask. The applications are labeled on the horizontal axis from 0 to 7. There is a box plot for each application. Each box shows the 25$^{th}$ and 75$^{th}$ percentile values for the application. The line inside the box is the median value. The whiskers of the box extend up to values not considered as statistical outliers. The statistical outliers are shown by a tail of `+' markers. Note that the Y axis on the three charts is different -- Figures~\ref{subfig_pcbox} and~\ref{subfig_pnsbox} have a magnified view of the values for legibility.

With the Noisy Baseline, the average trace of each application has a distinct distribution of values, leaving a fingerprint of that application. With a Constant mask, the median values of the applications are closer than they were with the Baseline. The lengths of the boxes do not differ as much as they did with the Baseline. Unfortunately, each application has a different statistical fingerprint sufficient to distinguish it from the others.  

Finally, with \des Gaussian Sinusoid mask, the distributions are {\em near-identical}.
This is because the patterns in each run are not correlated with those in other runs. Therefore, averaging multiple \des signals cancels out the patterns --- simply leaving a constant-like value with small variance. Hence, the median, mean, variance, and the distribution of the samples are very close. Note that the resolution of this data is 0.01W, indicating a high degree of obfuscation.

To highlight the differences further, Figure~\ref{fig_avg} shows the averaged signals of three applications over timesteps for the Noisy Baseline, \des Constant and \des Gaussian Sinusoid. Again, the Y axis for Figures~\ref{subfig_pcav} and~\ref{subfig_pnsav} is magnified for legibility. With Noisy Baseline, 
after the noise is removed by the averaging effect, the patterns in the averaged signals of each application are clearly visible. Further, these patterns are different for each application.

When the Constant mask is used, the magnitude of the variation is lower, but the
lines are not identical across applications. The pattern of blackscholes is clearly visible.

With the Gaussian Sinusoid mask, the average signal of an application has only a small variance, and is close to the average signals of the other applications. As a result, the average traces of different applications are indistinguishable, and do not resemble the baseline at all. This results in the highest degree of obfuscation.  

\begin{figure*}[htb]
\centering
\subfloat{
\includegraphics[width=0.26\textwidth]{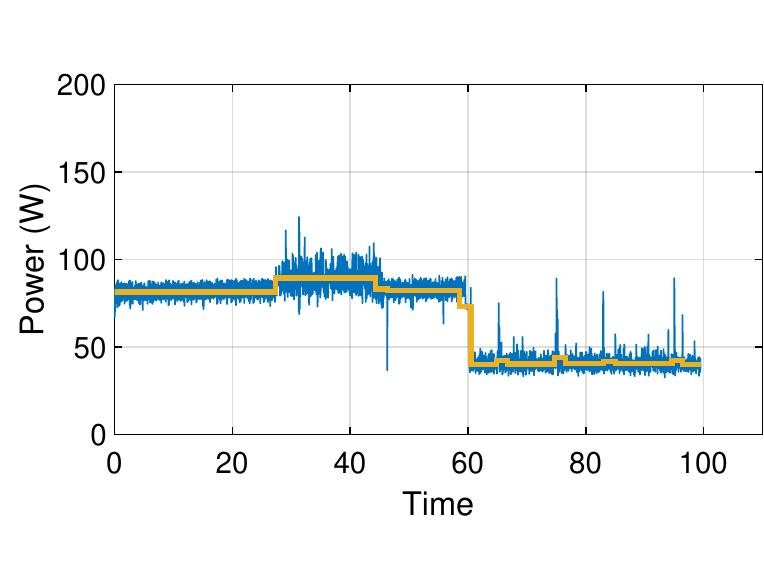}
\label{subfig_bls_baseline}
}%
\subfloat{
\includegraphics[width=0.26\textwidth]{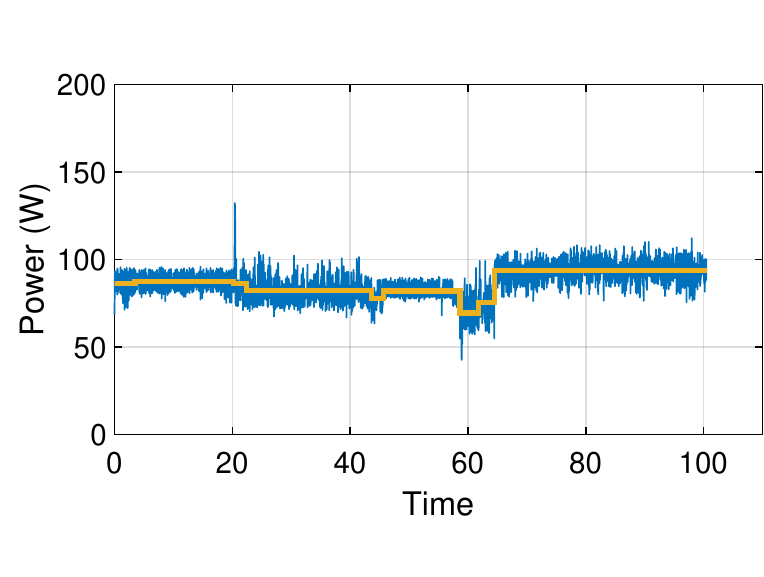}
\label{subfig_bls_constant}
}%
\subfloat{
\includegraphics[width=0.26\textwidth]{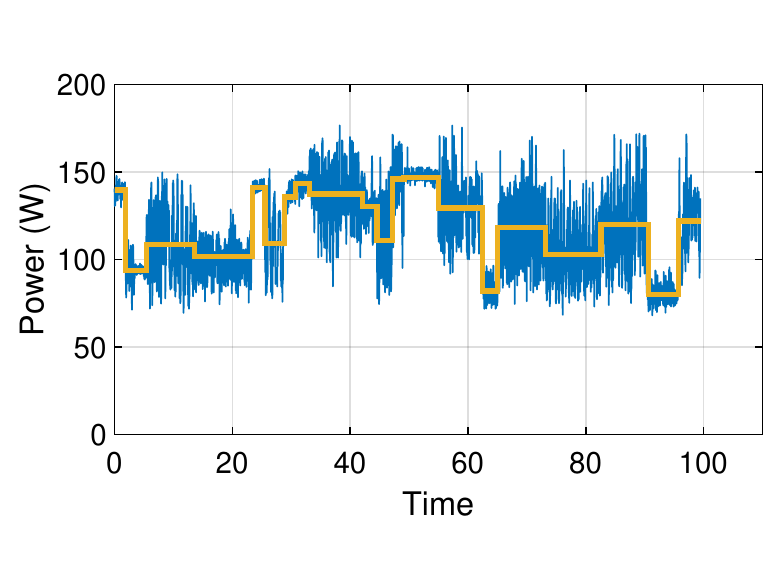}
\label{subfig_normalsine}
}\vspace{-5mm}\\
\setcounter{subfigure}{0}
\subfloat[Noisy Baseline]{
\includegraphics[width=0.2\textwidth]{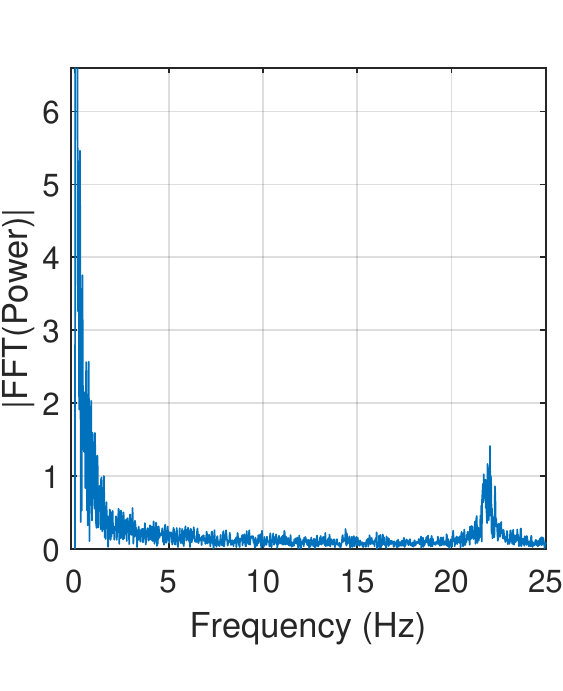}
\label{subfig_bl_fftb}
}\hspace{1cm}
\subfloat[\des Constant]{
\includegraphics[width=0.2\textwidth]{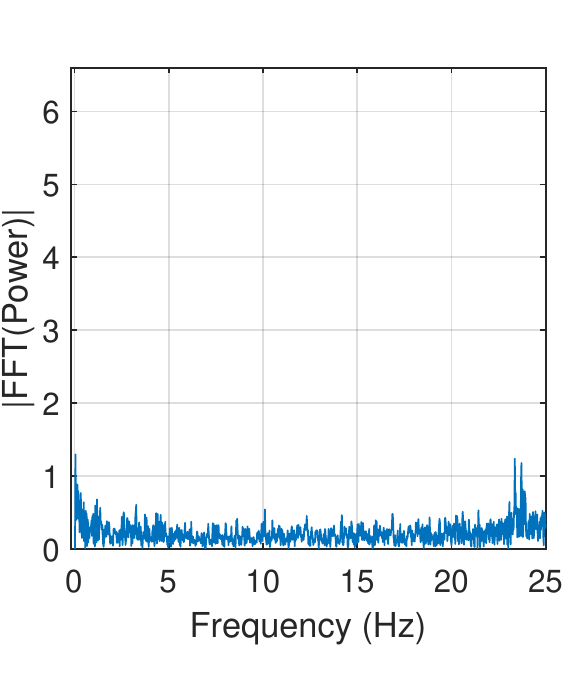}
\label{subfig_bl_fftc}
}\hspace{1cm}
\subfloat[\des Gaussian Sinusoid]{
\includegraphics[width=0.2\textwidth]{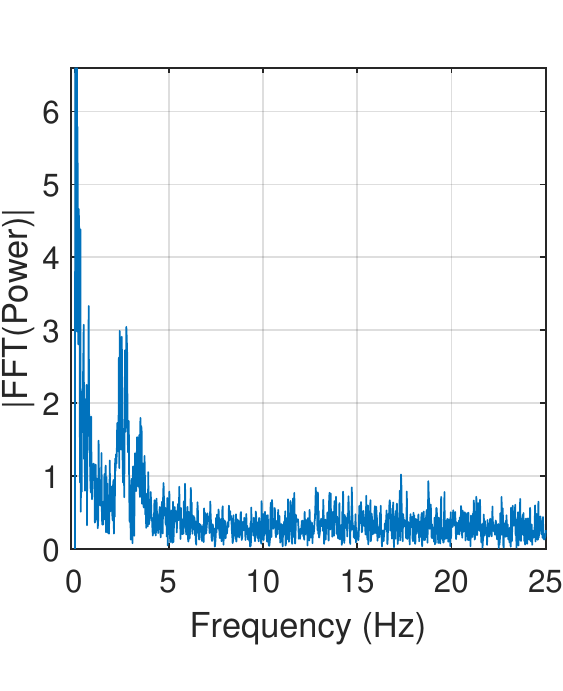}
\label{subfig_bl_fftns}
}%
\caption{Change point detection in blackscholes on \textit{Sys2}. The top row shows 
the time-domain signals, along with the detected phases. The bottom row shows the frequency-domain signals.\label{fig_bl}}
\vspace{-4mm}
\end{figure*}

\noindent
\textbf{Change Point Detection:} We present the highlights of this analysis using \textit{blackscholes} on \textit{Sys2}. We monitor the execution for 100 s,  even if the application completes before 100 s. Figure~\ref{fig_bl} shows the power signals 
in the time and frequency domains for blackscholes with the Noisy Baseline, \des Constant, and \des Gaussian Sinusoid. The time-domain plots also show the detected phases from the change point algorithm. 

In the Noisy Baseline (Figure~\ref{subfig_bl_fftb}),  the difference between the different phases is less visible due to interference from the idle threads. Nonetheless, the algorithm detects four major phases. They correspond to the sequential, parallel, sequential, and fully idle activity, respectively. The sudden changes between phases can be seen as a small spike in the FFT tail of this signal.

Figure~\ref{subfig_bl_fftc} shows blackscholes with a Constant mask. Change point analysis reveals changes in the signal at 20s, 40s and 60s, yielding four phases.  These phases can be related directly to those in the Baseline signal because the Constant mask does not introduce artificial changes. The change in signal variance between these phases is visible from the time series. Also,
the FFT tail has a small spike at the same location as with the Baseline. 
As a result, the attacker can easily identify this signal. 

Figure~\ref{subfig_bl_fftns} shows the behavior with the Gaussian Sinusoid
mask. Change point analysis detects several instances of phase change, {\em but these are all artificial}. Notice that the FFT of this signal is different from the Baseline FFT, eliminating any identity of the application. Finally, it is also {\em not possible to infer when the application execution is complete}. Specifically, the application actually completed around 55 s, but the power signal has no distinguishing change at that time.



\subsection{Effectiveness of a Control-Theory Controller}

\begin{figure}[h!]
\centering
\subfloat[Ideal target power given by the mask.]{
\includegraphics[width=0.7\columnwidth]{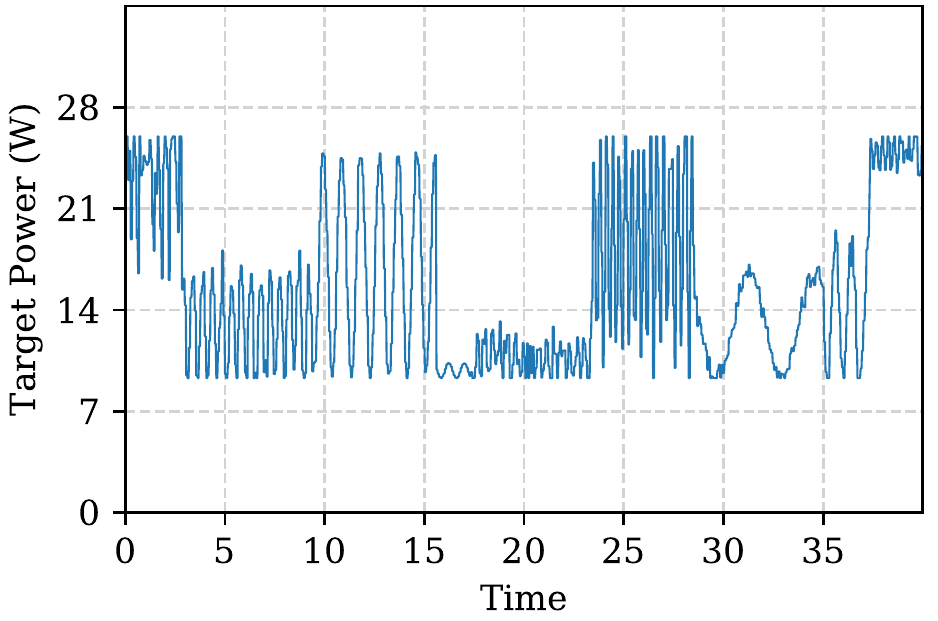}
\label{subfig_pmeas}
}
\\
\subfloat[Measured power from the system.]{
\includegraphics[width=0.7\columnwidth]{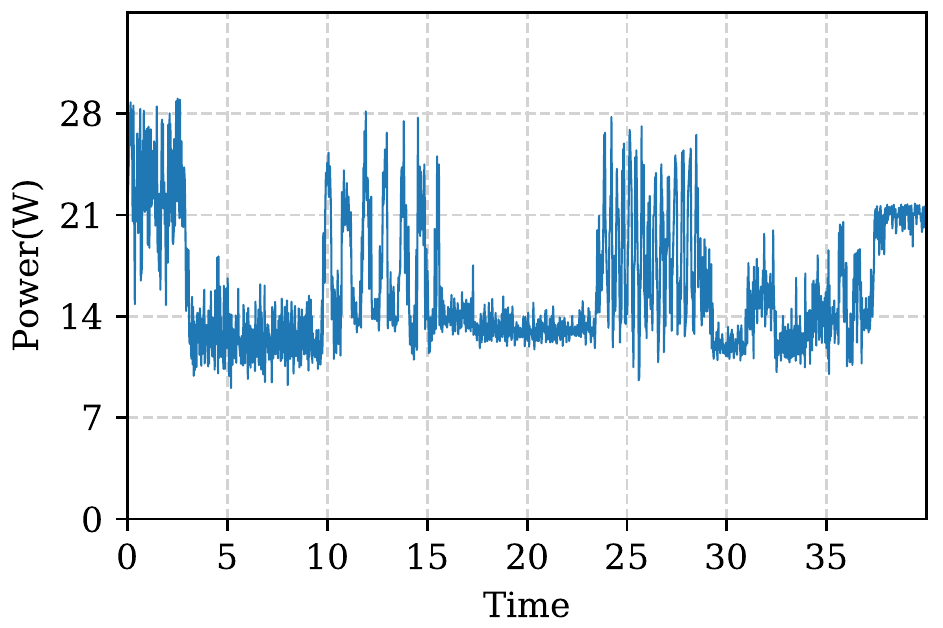}
\label{subfig_ptarg}
}
\caption{Target and measured powers for a run of blackscholes with \des Gaussian Sinusoid on \textit{Sys1}, showing the high-fidelity power-shaping with control theory.\label{fig_targ}}
\end{figure}

Figure~\ref{fig_targ} shows the target power given by a \des Gaussian Sinusoid mask generator during one execution of blackscholes on \textit{Sys1} (a), and the actual power that was measured from the computer (b). It can be seen that the 
control-theory controller is effective at making the measured power appear 
close to the target mask. This is thanks to the advantages of using a MIMO control theoretic controller. Indeed, this accurate tracking is what helps \des in effectively re-shaping the system's power and hide application activity.

\vspace{-1mm}

\subsection{Overheads and Power-Performance Impact}
\label{sub_overheads}

Finally, we examine the overheads and the power-performance impact of \des.

\noindent
\textbf{Overheads of \des:} The controller reads one output, sets three inputs
and, it can be shown, has a state vector $x(T)$ that is 11-element long
(Equation~\ref{eq:control}). Therefore, the controller needs less than 1 KB of
storage. At each invocation, it performs $\approx$200 fixed-point operations to make a decision. This completes within one $\mu$s. 

The Mask Generator requires (pseudo) random numbers from a Gaussian distribution to compute the mask (Equation~\ref{eq_ns}). It also needs another set of random numbers to set the properties of the Gaussian distribution and the Sinusoid. In our 
implementation, we use a software library that takes less than 10 $\mu$s to generate all our random numbers. For a hardware implementation, there are off-the-shelf hardware instructions and IP modules to obtain these random numbers in sub-$\mu$s~\cite{inteldrng,intelrandIp}.

\des needs few resources to control the system, making \des attractive for both hardware and software implementations. The primary bottlenecks in our implementation were the sensing and actuation latencies, which are in the ms time scale.

\noindent
\textbf{Application-Level Impact:} We run the PARSEC and SPLASH2x applications on \textit{Sys1} and \textit{Sys2}, with and without \des, to measure the power and performance overheads. Our Baseline configuration runs at the highest available frequency without interference from the idle threads or the balloon program. It offers no security. We also evaluate the  Noisy Baseline design, in which the system runs with a random DVFS level and percentage of idle activity. 

\begin{figure}[h]
\centering
\subfloat[Power]{
\includegraphics[width=0.9\columnwidth]{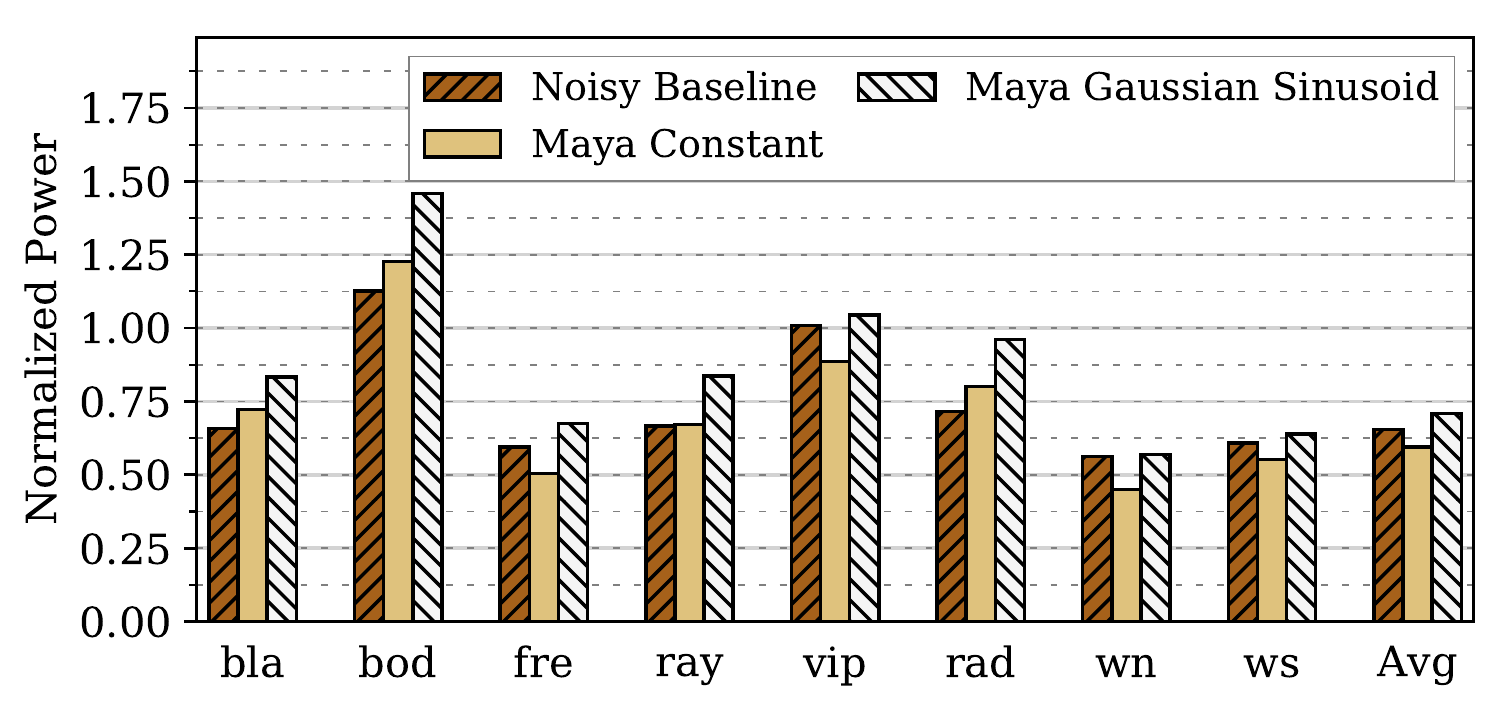}
\label{subfig_pp}
}\\ \vspace{-2mm}
\subfloat[Execution Time]{
\includegraphics[width=0.9\columnwidth]{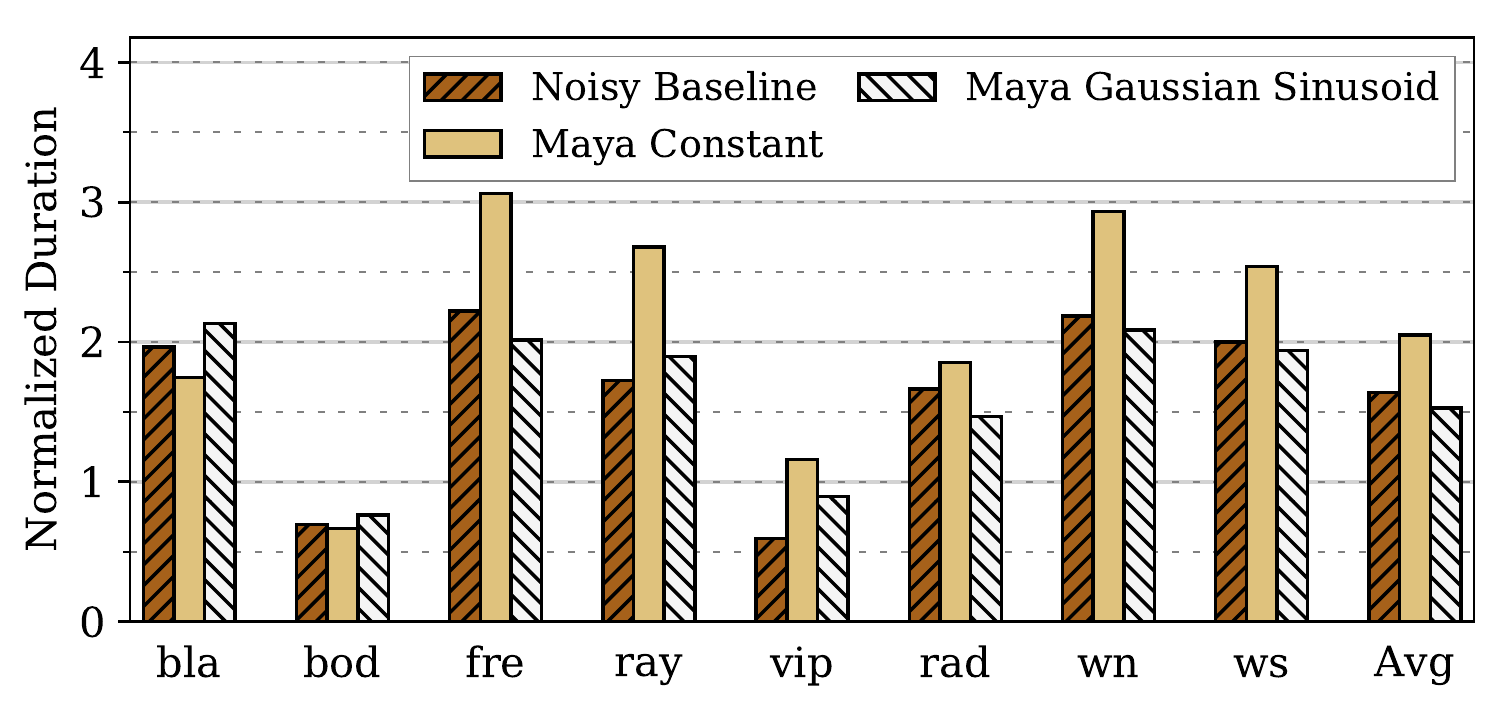}
\label{subfig_pt}
}
\vspace{-2mm}
\caption{Overheads of our environments on \textit{Sys1} relative to a high-performance insecure Baseline.\label{fig_pover}}
\vspace{-2mm}
\end{figure}

Figure~\ref{fig_pover} shows the power and execution time of the Noisy Baseline, \des Constant and \des Gaussian Sinusoid environments
normalized to that of the high-performance Baseline. From Figure~\ref{subfig_pp}, 
we see that the average power consumed by 
the applications with the three environments is 35\%, 41\%, and 29\% lower than that of the high performance Baseline, respectively. The power is lower 
due to the idle threads and low DVFS values that appear in these environments. 

Figure~\ref{subfig_pt} shows the normalized execution times with the Noisy Baseline, \des Constant, and \des Gaussian Sinusoid environments. These environments have, on average, performance slowdowns of 63\%, 100\% and 50\% respectively, over the Baseline. \des Constant uses a single power target that is lower than the maximum power at which Baseline runs. Therefore, its performance is the worst.  On the other hand, \des Gaussian Sinusoid spans multiple power levels in the available range, allowing applications to run at higher power occasionally. As a result, execution times are relatively better. \des Gaussian Sinusoid also has a better performance than Noisy Baseline and offers high security. 



It can be shown that the power and performance overheads of our environments
in \textit{Sys2} are similar to those in \textit{Sys1} shown in 
Figure~\ref{fig_pover}. This shows that our methodology is robust across
different machine configurations.

One approach to reduce the slowdown from \des is to enable obfuscation only on demand. Authenticated users or secure applications can activate \des before commencing a secure-sensitive task, and stop \des once the task is complete. While this approach gives away the information that a secure task is being run, at least does
not  slow down all applications.

\subsection{Overall Remarks}

Using machine learning-based attacks and signal analysis, we showed how the
\des Gaussian Sinusoid mask can obfuscate power signals from a computer. 
\des's security comes from an effective mask, and from
the control-theoretic controller 
that can shape the computer's power into the given mask. 
We implemented \des on two different machines \textit{without modifying} the controller or the mask generator, demonstrating  our proposal's robustness, security and ease of deployment on existing computers.

\section{Related Work}
\label{relwork}

Power, temperature, and EM emissions from a computer are a set of physical side channels that have been used by many attackers to compromise security~\cite{Kocher2011,sideChannelRetrospect,powerSide}. 
Kocher et al.~\cite{Kocher2011}  give a detailed overview of attacks exploiting power signals with Simple Power analysis (SPA) and Differential Power Analysis (DPA).  

Machine Learning (ML) is commonly-used to perform power analysis
attacks. Using ML, Yan et al. developed an attack to identify the running application, login screens and the number of keystrokes typed by the user~\cite{mobilePowerside}. Lifshits et al. considered malicious smart batteries and demonstrated another attack that predicted the character of a keystroke and user's personal information such as browser, camera, and location activity~\cite{batteryAttack}. Yang et al. snooped the power drawn by a mobile phone from a public USB charging booth to predict the user's browser activity. Hlavacs et al. showed that the virtual machines
running on a server could be identified using power signals~\cite{vmEnergyAttack}. 
Finally, Wei et al. used ML to detect malware from power signals~\cite{wei19host}.

Other attacks are even more sophisticated.
Michalevsky et al. developed a malicious smartphone application that could track the user's location using OS-level power counters, without reading the GPS module~\cite{powerspy}. 
Schellenberg et al. showed how malicious on-board modules can measure another chip's power activity~\cite{interChipPowerSide}.

As power, temperature, and EM emissions are correlated, attackers used temperature and EM measurements to identify application 
activity~\cite{genkinEm,GenkinLaptop,emAttack,emPower,emSmartCard, iotDeepLearnEm}. These attacks have targeted many environments, like smart cards, mobile systems, 
laptops, and IoT devices.
Recently, Masti et al. showed how temperature can be used to identify another core's activity in multicores~\cite{thermalMulticoreCovert}. 
This broadens the threat from physical side channels because attackers can read EM signals  from a distance, or measure temperature through co-location, even when the system does not support per-core power counters.

Several countermeasures against power side channels have been proposed~\cite{powerSide,constPowerLogic,maskDesEnergy,yang2005power,circuitPowerDefense,Kocher2011,dvfsDefense,randomDvsDfs}. They usually operate along one of 
two principles: keep power consumption invariant to any activity~\cite{constPowerLogic,maskDesEnergy, signalSuppress,onChipVrDefense,circuitPowerDefense}, or make power consumption noisy such that the impact of application activity is lost~\cite{Kocher2011,powerSide}. A common approach is to randomize DVFS  using special hardware~\cite{dvfsDefense,randomDvsDfs,dvfsDefense}. Avirneni and Somani also propose new circuits for randomizing DVFS, and change voltage and frequency independently~\cite{randomDvsDfs}. 

Baddam and Zwolinski showed that randomizing DVFS is not a viable defense because attackers can identify clock frequency changes through high-resolution power traces~\cite{dvfsSecurityEval}. Yang et al. proposed using random task scheduling at the OS level in addition to new hardware for randomly setting the processor frequency and clock phase~\cite{multicorePowerSideCounter}. 
Real et al. showed that simple approaches to introduce noise or empty activity into the system can be filtered out~\cite{shapeSignal}.

Trusted execution environments like Intel SGX~\cite{sgx} or ARM Trustzone~\cite{trustzone} can sandbox the architectural events of applications. However, they do not establish boundaries for physical signals. One approach for power side channel isolation is blinking~\cite{blinking}, where a circuit is temporarily cut-off from the outside and is run with a small amount of energy stored inside itself~\cite{blinking}. 

To the best of our knowledge, \des is the first 
defense against power side channels that uses control theory.


\vspace{-2mm}
\section{Conclusions}
\label{conclude}

This paper presented a simple and effective solution against power
side channels. The idea, called {\em \des},
is to use a controller from control theory 
to distort, in an application-transparent way, the 
power consumed by an application -- so that the attacker 
cannot obtain information.
With control theory techniques, the controller can  
keep outputs close
to desired targets even when runtime conditions change
unpredictably.  Then, by changing these targets 
appropriately, \des makes the power signal
appear to carry activity information which, in
reality, is unrelated to the program.
\des controllers can be implemented in privileged
software or in simple hardware.  In this paper, we 
implemented \des on two multiprocessor machines using OS
threads, and showed that
it is very effective at falsifying application activity.

\bibliographystyle{ACM-Reference-Format}
\bibliography{IEEEabrv,references}

\end{document}